\documentclass[12pt]{article}

\def\lromn#1{\uppercase\expandafter{\romannumeral#1}}
\def\ds{\displaystyle}
\def\vev#1{\langle 0 |#1|0 \rangle}

\def\Slash#1{{\ooalign{\hfil$#1$\hfil\crcr\hfil$/$\hfil}}}
\def\SlashIt#1{{\ooalign{\hfil$#1$\hfil\crcr\hfil{\it/}\hfil}}}
\newcommand{\gsim}{ \mathop{}_{\textstyle \sim}^{\textstyle >} }
\newcommand{\lsim}{ \mathop{}_{\textstyle \sim}^{\textstyle <} }

\usepackage{graphicx}
\usepackage{latexsym}
\usepackage{amssymb}
\usepackage{amsmath}

\begin{document}

\begin{flushright}
 ICRR-Report-513-2004-11\\
 YITP-04-73\\
 \today\\
\end{flushright}

\begin{center}

  \begin{large}
    \textbf{Non-Perturbative Effect on Dark Matter Annihilation\\
   and Gamma Ray Signature from Galactic Center}
  \end{large}

  \vspace{2cm}

  \begin{large}
    Junji Hisano$~^a$, Shigeki. Matsumoto$~^a$, Mihoko M. Nojiri$~^b$,
   and Osamu Saito$~^a$

    $~^a$
    ICRR,
    University of Tokyo,
    Kashiwa 277-8582,
    Japan
    \\
    $~^b$
    YITP, 
    Kyoto University, 
    Kyoto 606-8502, 
    Japan
  \end{large}

  \vspace{3cm}

  {\bf ABSTRACT}

\end{center}

Detection of gamma rays from dark matter annihilation in the
galactic center is one of the feasible techniques to search for 
dark matter. We evaluate the gamma ray flux in the case that the dark
matter has an electroweak SU(2)$_L$ charge. Such dark matter is
realized in the minimal supersymmetric standard model (MSSM) when the
lightest SUSY particle is the Higgsino- or Wino-like neutralino. When the
dark matter is heavy compared to the weak gauge bosons, the
leading-order calculation of the annihilation cross sections in
perturbation breaks down due to a threshold singularity. We take
into account non-perturbative effects by using the non-relativistic
effective theory for the two-body states of the dark matter and its
SU(2)$_L$ partner(s), and evaluate precise cross sections
relevant to the gamma ray fluxes. We find that the annihilation cross
sections may be enhanced by several orders of magnitude due to 
resonances when the dark matter mass is larger than 1
TeV. Furthermore, the annihilation cross sections in the MSSM may be
changed by factors even when the mass is about 500 GeV. We also
discuss sensitivities to gamma ray signals from the galactic
center in the GLAST satellite detector and the large Air Cerenkov
Telescope arrays.

\newpage


\vspace{1.0cm}
\lromn 1 \hspace{0.2cm}
{\bf Introduction}
\vspace{0.5cm}

Recent cosmological observations determine precisely the mean densities
of matter and baryon in the Universe \cite{Spergel:2003cb},
and existence of non-baryonic dark matter is established. Weakly
interacting massive particles (WIMPs) are considered to be good candidates
of the dark matter \cite{Jungman:1995df}. They act as the cold dark matter
in the structure formation of the universe. High resolution $N$-body
simulations show that the cold dark matter hypothesis explains well the
structure larger than about 1 Mpc \cite{Primack:2002th}. On the other
hand, fundamental problems, such as (i) the constituent of the dark
matter  and the origin in the thermal history and (ii) the dark matter
distribution in the galactic scale, are not still resolved. It is
important to detect the dark matter in direct or indirect methods in
order to answer the questions. 

Many detection methods have been proposed and some of the experiments
are now operating. Among those, the detections of exotic cosmic ray
fluxes, such as positrons, anti-protons and gamma rays, are  feasible
techniques to search for the dark matter particles
\cite{Silk:1985ax}-\cite{Bergstrom:1997fj}. In particular, an excess of   
monochromatic (line) gamma rays due to the pair annihilation would be a
robust signal if observed, because diffused gamma ray background
must have a continuous energy spectrum \cite{Bergstrom:1997fj}. The
GLAST satellite detector \cite{Morselli:1997rk} and the large
Atmospheric Cerenkov Telescope (ACT) arrays, such as CANGAROO III
\cite{Tsuchiya:2004wv}, HESS \cite{Hinton:2004eu}, MAGIC
\cite{Baixeras:2003xr} and VERITAS \cite{Weekes:1997np}, can search for
the exotic gamma rays from the galactic center, the galactic halo, and
even from extra galaxies.

In this paper, we discuss SU(2)$_L$ non-singlet WIMPs and the gamma ray
fluxes from the galactic center due to the pair annihilation. We refer
to such dark matter  as electroweak-interacting massive particle
(EWIMP) dark matter \cite{Hisano:2002fk}. Due to their SU(2)$_L$
non-singlet nature, EWIMPs have interactions with the SU(2)$_L$
gauge bosons such as $W$ and $Z$ bosons. If EWIMPs have a vector
coupling to $Z$ boson, the current bound obtained from direct dark
matter searches through their spin-independent interaction is stringent
\cite{Akerib:2003px}. This means that the EWIMP dark matter should be a
Majorana fermion or a real scalar if the mass is around 1 TeV. In this
paper, we consider the former 
case. We especially study triplet and doublet EWIMP dark matters, which
are neutral components of an SU(2)$_L$-triplet fermion whose
hypercharge is zero and of a pair of SU(2)$_L$-doublet fermions with
hypercharges $\pm 1/2$, respectively. When the EWIMP mass is around 1
TeV, the thermal relic abundance may be consistent with the
cosmological observation.

The EWIMP is realized in the minimal supersymmetric standard model
(MSSM) when the lightest SUSY particle (LSP) is the Higgsino- or
Wino-like neutralino \cite{Haber:1984rc}. Wino is the superpartner of
the SU(2)$_L$ gauge boson, and Higgsino is that of SU(2)$_L$ doublet
Higgs bosons. The thermal relic density is too low if
the LSP mass is smaller than 1 TeV.  However, decays of gravitino or
other quasi-stable particles may produce the LSPs non-thermally so
that the relic abundance is consistent with the cosmological
observation. While the LSP with the mass about 1 TeV may lead to the
naturalness problem, such possibilities are discussed in the split
SUSY scenario \cite{Arkani-Hamed:2004fb}.

The line and continuum gamma ray fluxes from the EWIMP dark matter
annihilation in the galactic center are proportional to the cross
section to two photons and those to other modes whose final states
fragment into $\pi^0$s, respectively. The leading-order cross sections
in perturbation have been calculated by many authors
\cite{Bergstrom:1997fh}. However, if the EWIMP mass is large compared to
the weak gauge boson masses, usual perturbative expansion for the
annihilation cross sections cannot be applied \cite{Hisano:2002fk}.
This can be seen in the violation of unitarity of the one-loop
annihilation cross section into two photons.  The fact comes from the
degeneracy of the EWIMP and its SU(2)$_L$ partner(s) in mass and the
non-relativistic motion of the dark matter in the current
universe. The transition between EWIMP and its partner pairs is
induced by the $t$-channel weak gauge  boson exchange. When the EWIMP
mass is much larger than the weak gauge boson masses, the weak interaction
behaves as a long-range force. The wave functions of EWIMP and its
partner pairs are modified from plane waves at the non-relativistic
limit, and the mixing between those states is enhanced. This
phenomenon is related to so-called a threshold singularity, and we
have to consider the effects of the long-range force on the
annihilation cross sections for reliable calculation.

In this paper we work in the non-relativistic effective theory for
EWIMPs. Non-relativistic effective theories
\cite{Caswell:1985ui} are often used in calculations of the threshold
productions of heavy particles, the quarkonium mass spectrums and so
on \cite{Strassler:1990nw}.  In this technique, we can factorize 
short-distance physics, such as pair annihilation, from 
long-range effects on the wave functions due to the optical theorem
\cite{Bodwin:1994jh}. The long-range effects are evaluated by solving the wave
functions under the potential.

We found that the annihilation cross sections may be enhanced by
several orders of magnitude compared to the leading-order calculation
in perturbation when the EWIMP mass is larger than about 1 TeV. 
The mixture of the pairs of EWIMPs and the
SU(2)$_L$ partners can form a bound state whose binding energy is close
to zero, and it contributes to the annihilation cross sections in the
non-relativistic limit. The enhancement of the cross sections
originates from the resonance by the bound state.  Furthermore, the
annihilation cross section to two photons, which is suppressed by a
loop factor in perturbation, becomes comparable to those to the other
modes around the resonance. As a result, the continuum and line gamma
ray fluxes from the galactic center due to the EWIMP annihilation are
enhanced. The indirect dark matter searches by the large ACT detectors,
which have sensitivities to TeV-scale  gamma rays, may be promising, if
dark matter is a TeV EWIMP.

It is also found that the non-perturbative corrections to the cross
sections are sizable for the triplet (doublet) EWIMP even when the
EWIMP mass is about 500 (1500) GeV. Thus, the correction should be taken
into account in the evaluation of the gamma ray fluxes in the MSSM,
especially when the LSP is Wino-like.

This paper is organized as follows. We first summarize the properties
of EWIMPs and discuss the threshold singularity in the EWIMP pair
annihilation in the next section. In Section III the non-relativistic
effective actions for the triplet and the doublet EWIMP pairs are
derived. In Section IV the cross section formula is obtained using the
optical theorem and the effective actions. While the obtained
annihilation cross section to two photons in our formula is reduced to
the one-loop result in the perturbative expansion, it also satisfies
the unitarity bound in the limit of an infinite EWIMP mass. The one-loop
cross section does not satisfies the bound.

In Section V some numerical results for the annihilation cross
sections are presented. The fitting formulae for the annihilation
cross sections are also derived from the numerical results. The
resonance behaviors of the cross sections are studied using a 
toy model in which the electroweak Yukawa potentials are approximated by
a well potential. In Section VI we evaluate the gamma ray fluxes from
the EWIMP annihilation in the galactic center and discuss the
sensitivities in the future experiments. In Sections V and VI the
cross sections and the gamma ray fluxes for the Wino- and the
Higgsino-like neutralinos are also evaluated in the wide range of the
MSSM parameters.  Section VII is devoted to summary of the paper.


\vspace{1.0cm}
\lromn 2 \hspace{0.2cm} {\bf Properties of EWIMPs and Threshold Singularity}
\vspace{0.5cm}

In this section the mass spectrums of EWIMPs and the SU(2)$_L$
partners and their low-energy interactions are summarized. We also
discuss the threshold singularity in the non-relativistic EWIMP pair
annihilation. The perturbative expansion of the annihilation cross
sections is spoiled due to the singularity when the EWIMP mass is
heavy compared to the weak gauge boson masses.

\vspace{0.5cm}
\underline{1. Properties of EWIMPs}
\vspace{0.5cm}

The EWIMP dark matter $\tilde{\chi}^0$ is a neutral component of
SU(2)$_L$ multiplet(s). In this paper, we consider two cases. One is an
SU(2)$_L$ triplet fermion whose hypercharge is zero. In this case, the
EWIMP is accompanied with the SU(2)$_L$ partner, a charged Dirac
fermion $\tilde{\chi}^-$. They are almost degenerate in mass, and the
mass difference $\delta m$ is caused by the electroweak symmetry
breaking. If $\delta m$ comes from the radiative correction of the  gauge
boson loops \cite{Cheng:1998hc},
\begin{eqnarray}
 \delta m_{\rm rad}
 &=&
 \frac{\alpha_2m}{4\pi}
 \left[
  f\left(\frac{m_W}{m}\right)
  -c_W^2f\left(\frac{m_Z}{m}\right)
  -s_W^2f(0)
 \right]~,
 \nonumber \\
 \nonumber \\
 f(a)
 &=&
 \int_0^1 dx 2(1 + x)\log[x^2 + (1 - x)a^2]~,
 \label{dmcor}
\end{eqnarray}
where $m$ is the EWIMP mass, $\alpha_2$ is the SU(2)$_L$ gauge
coupling, $m_W(m_Z)$ is the $W(Z)$ gauge boson mass, and $c_W(s_W) =
\cos\theta_W$$(\sin\theta_W)$, where $\theta_W$ is the Weinberg
angle. The gauge interactions of the EWIMP and its SU(2)$_L$ partner
are given by
\begin{eqnarray}
 {\cal L}_{\rm int}
 &=&
 -\frac{e}{s_W}
 \left(
  \overline{\tilde{\chi}^0}\Slash{W}^\dagger\tilde{\chi}^-
  +
  {\rm h.c.}
 \right)
 +
 e \frac{c_W}{s_W}\overline{\tilde{\chi}^-}\Slash{Z}\tilde{\chi}^-
 +
 e\overline{\tilde{\chi}^-}\Slash{A}\tilde{\chi}^-~,
 \label{gauge interaction of Wino}
\end{eqnarray}
where $e = \sqrt{4\pi\alpha}$ and $\alpha$ is the fine structure
constant. The mass difference $\delta m_{\rm rad}$ is induced by the
custodial SU(2)$_L$ symmetry breaking in the gauge sector, and $\delta
m_{\rm rad} \simeq 0.18$ GeV if $m \gg m_W$ and $m_Z$. Effective
higher-dimensional operators may also generate the mass difference,
however it is suppressed by $m_W^4/\Lambda^3$, where $\Lambda$ is a new
particle mass scale.

Another example of the EWIMP dark matter is a neutral component in a
pair of SU(2)$_L$ doublet fermions with the hypercharges $\pm
1/2$. After the symmetry breaking, two neutral mass eigenstates,
$\tilde{\chi}^0$ and $\tilde{\chi}^0_N$, appear. The lightest one is a 
candidate of the EWIMP dark matter. A charged Dirac fermion
$\tilde{\chi}^-$ is also accompanied with them, and they are also
degenerate in mass in the SU(2)$_L$ symmetric limit. The mass
differences among them are generated by effective operators via the 
electroweak symmetry breaking. Unlike the triplet EWIMP case, the mass
difference is ${\cal O}(m_W^2/\Lambda)$ and it is not strongly
suppressed by $\Lambda$. The gauge interactions of the doublet EWIMP
dark matter and its partners are given by 
\begin{eqnarray}
  {\cal L}_{\rm int}
  &=&
  -\frac{e}{2s_W}
  \left(
    \overline{\tilde{\chi}^0}\Slash{W}^\dagger\tilde{\chi}^-
    -
    \overline{\tilde{\chi}^0_N}\Slash{W}^\dagger\tilde{\chi}^-
    +
    {\rm h.c.}
  \right)
  -
  \frac{e}{s_Wc_W} \left(\frac{1}{2} - c_W^2\right)
  \overline{\tilde{\chi}^-}\Slash{Z}\tilde{\chi}^-
  \nonumber \\
  &~&
  +
  e\overline{\tilde{\chi}^-}\Slash{A}\tilde{\chi}^-
  +
  \frac{e}{2s_Wc_W}
  \overline{\tilde{\chi}^0}\Slash{Z}\tilde{\chi}^0_N~.
\end{eqnarray}

An example of the EWIMP dark matter is the lightest neutralino in the
MSSM. Neutralinos $\tilde{\chi}^0_i$ ($i = 1\cdots 4$) are linear
combinations of the supersymmetric partners of gauge bosons and Higgs
bosons, Bino ($\tilde{B}$), neutral Wino ($\tilde{W}^0$) and neutral
Higgsinos ($\tilde{H_1^0}$, $\tilde{H_2^0}$). While those four fields
have SU(2)$_L \otimes$ U(1)$_Y$ invariant masses, they are mixed with
each other via the electroweak symmetry breaking \cite{Haber:1984rc},
\begin{eqnarray}
 \tilde{\chi}^0_i
 =
   Z_{i1}\tilde{B}
 + Z_{i2}\tilde{W}^0
 + Z_{i3}\tilde{H}_1^0
 + Z_{i4}\tilde{H}_2^0~.
 \label{neutralino comp}
\end{eqnarray}
Coefficients $Z_{ij}$ are determined by diagonalizing the neutralino
mass matrix,
\begin{eqnarray}
 M_{\tilde{\chi}^0} = 
 \left(
  \begin{array}{cccc}
   M_1 & 0 & -m_Z~s_W~c_\beta & m_Z~s_W~s_\beta
    \\
   0 & M_2 & m_Z~c_W~c_\beta & -m_Z~c_W~s_\beta
    \\
   -m_Z~s_W~c_\beta & m_Z~c_W~c_\beta & 0 & -\mu
    \\
   m_Z~s_W~s_\beta & -m_Z~c_W~s_\beta & -\mu & 0
    \\
  \end{array}
 \right)~,
 \label{neutmass}
\end{eqnarray}
which is written in the $(\tilde{B}, \tilde{W}^0, \tilde{H_1^0}$,
$\tilde{H_2^0})$ basis. Here $M_1$ and $M_2$ are the Bino and Wino
masses, respectively, and $\mu$ is the supersymmetric Higgsino mass. The
variable $\tan\beta$ is given by the ratio of the vacuum expectation values
of two Higgs fields, and $c_\beta = \cos\beta$ and $s_\beta =
\sin\beta$. The lightest neutralino is Wino-like when $M_2 \ll |\mu|,
M_1$, and Higgsino-like when $|\mu| \ll M_1, M_2$. These two neutralinos
have SU(2)$_L$ charges and are candidates of the EWIMP dark matter; the
Wino-like neutralino is a triplet EWIMP and the Higgsino-like neutralino
is a doublet EWIMP.

Neutralinos are accompanied with charginos $\chi^-_i$ ($i = 1,
2$), which are linear combinations of charged Wino $\tilde{W}^-$ and 
charged Higgsino $\tilde{H}^- = \tilde{H}_{1L}^- +
\tilde{H}_{2R}^-$ \cite{Haber:1984rc}. The compositions of charginos are
determined by diagonalizing the chargino mass matrix,
\begin{eqnarray}
 M_{\chi^\pm} = 
 \left(
  \begin{array}{cc}
   M_2 & \sqrt{2}~m_W~s_\beta
    \\
   \sqrt{2}~m_W~c_\beta & \mu
    \\
  \end{array}
 \right)~,
 \label{chargmass}
\end{eqnarray}
which is written in the $(\tilde{W}^-,\tilde{H}^-)$ basis.

From the above matrices in Eqs. (\ref{neutmass}) and (\ref{chargmass}),
the mass difference $\delta m_{\rm tree}$ between the lightest
neutralino and  chargino at tree level can be calculated. If the
LSP is Wino-like ($m_Z,M_2\ll M_1,|\mu|$), $\delta m_{\rm tree}$
is approximately given by
\begin{eqnarray}
 \delta m_{\rm tree}
 \simeq
 \frac{m_Z^4}{M_1\mu^2}s_W^2c_W^2\sin^22\beta~,
\end{eqnarray}
which is suppressed by the third power of the high energy scale
$M_1\mu^2$ as discussed before. Since their masses are highly degenerate
at tree level, the radiative correction to the mass difference in
Eq.~(\ref{dmcor}) is also important.

The mass splitting for the Higgsino-like LSP in a case with $m_Z,|\mu| \ll M_1,M_2$
is ${\cal O}(m_Z^2/m_{\rm SUSY})$ and given by 
\begin{eqnarray}
 \delta m
 \simeq
 \frac{1}{2}\frac{m_Z^2}{M_2}c_W^2(1 - \sin 2\beta)
 +
 \frac{1}{2}\frac{m_Z^2}{M_1}s_W^2(1 + \sin 2\beta)~.
\end{eqnarray}
The second lightest neutralino is also degenerate with the LSP and the
chargino in mass, because they are in common SU(2)$_L$ multiplets. The 
mass difference $\delta m_N$  between the LSP and the second lightest
neutralino is again ${\cal O}(m_Z^2/m_{\rm SUSY})$,
\begin{eqnarray}
 \delta m_N
 \simeq
 \frac{m_Z^2}{M_2}c_W^2
 +
 \frac{m_Z^2}{M_1}s_W^2~,
\end{eqnarray}
and this is $2\times \delta m$ when $\tan \beta \gg 1$.

\begin{figure}[t]
 \begin{center}
  \includegraphics[height = 8.5cm,clip]{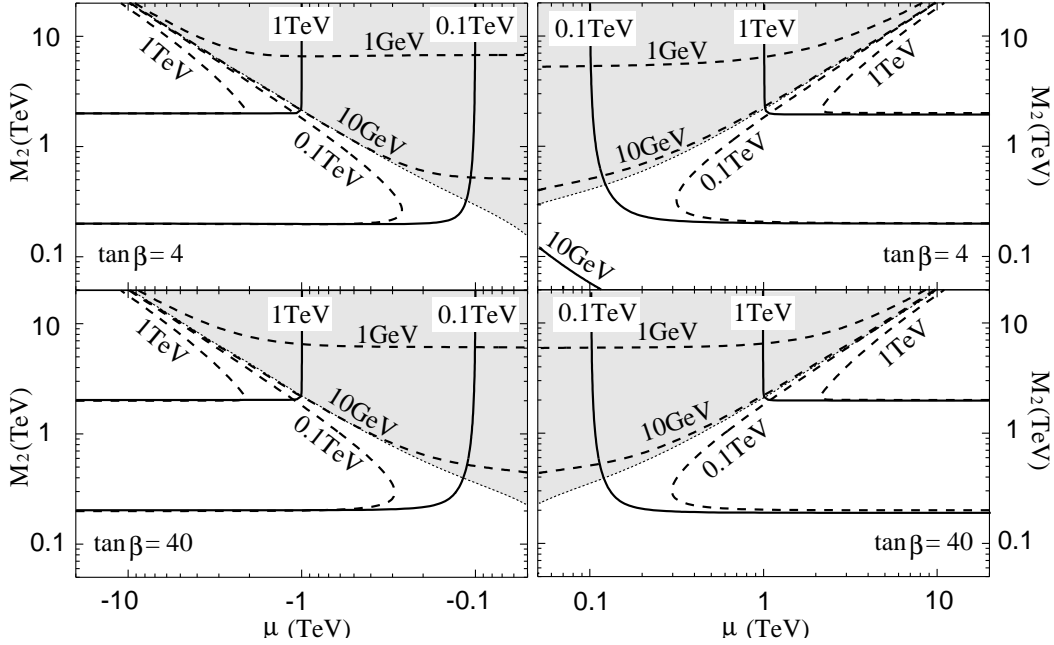}
 \end{center}
 \caption{\small
 Contour maps of the lightest neutralino mass (solid line) and the mass
 difference between the lightest neutralino and chargino (dashed
 line) in $(M_2, \mu)$ planes with $\tan\beta = 4$ (two top figures) and
 $\tan\beta = 40$ (two bottom figures) in the MSSM. $M_2 = 2M_1$ 
 is assumed. Shaded areas correspond to the Higgsino-like region
 ($|Z_{13}|^2 + |Z_{14}|^2 > 0.9$).
 \label{massdiff1}
 }
\end{figure}

\begin{figure}[t]
 \begin{center}
  \includegraphics[height = 8.5cm,clip]{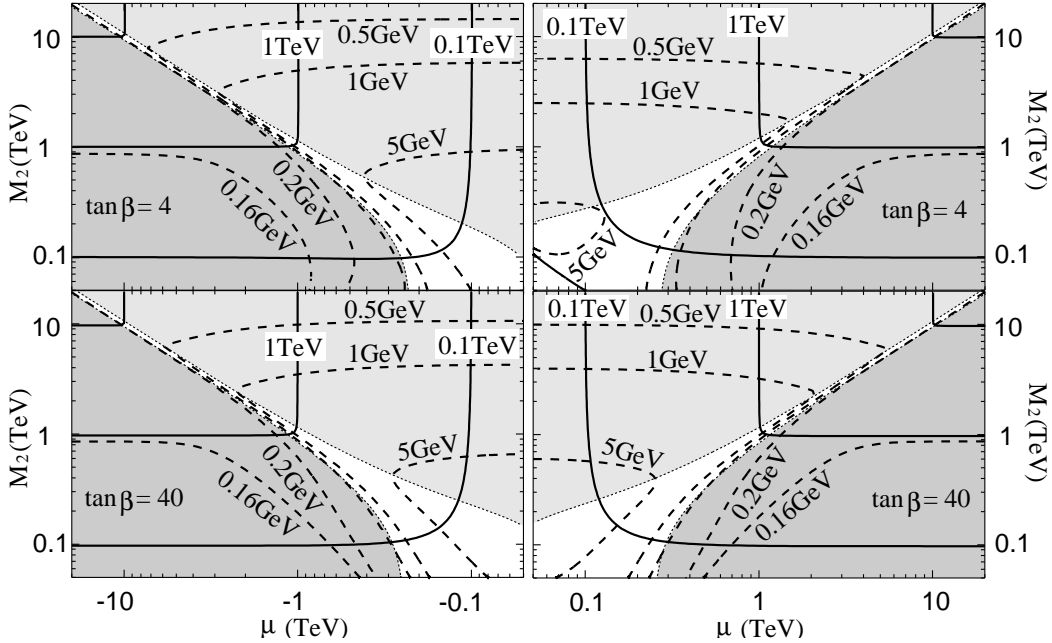}
 \end{center}
 \caption{\small
 Contour maps of the lightest neutralino mass (solid line) and the mass
 difference between the lightest neutralino and  chargino (dashed
 line) in $(M_2, \mu)$ planes with $\tan\beta = 4$ (two top figures) and
 $\tan\beta = 40$ (two bottom figures) in the MSSM. $M_2 = M_1/3$ 
 is assumed. Lighter shaded areas correspond to the Higgsino-like region
 ($|Z_{13}|^2 + |Z_{14}|^2 > 0.9$), while the darker shaded areas are
 the Wino-like one ($|Z_{12}|^2 > 0.9$). The radiative correction in
 Eq.~(\ref{dmcor}) is included for depicting the contour of  the mass
 difference in the Wino-like regions.
 \label{massdiff2}
 }
\end{figure}

In Fig.~\ref{massdiff1} and Fig.~{\ref{massdiff2}}, we show contours
of the lightest neutralino mass and the mass difference between the
neutralino and the lightest chargino in $(\mu, M_2)$ planes with
$\tan\beta = 4, 40$. These figures are obtained by diagonalizing the
mass matrices in Eqs. (\ref{neutmass}) and (\ref{chargmass})
numerically. In Fig.~\ref{massdiff1}, we assume the GUT relation
between the gaugino masses, $M_2 = 2M_1$. In this case, the
Higgsino-like neutralino (doublet EWIMP) may be the dark matter if
$|\mu| \lsim M_1$. The shaded areas in these figures correspond to the
Higgsino-like region ($|Z_{13}|^2 + |Z_{14}|^2 > 0.9$), and the
lightest neutralino is degenerate with the lightest chargino,
especially at a large mass.

Fig.~\ref{massdiff2} is the same plots except that we assume the
relation $M_2 = M_1/3$, which is predicted in the anomaly mediated
supersymmetry breaking scenario \cite{Randall:1998uk}. In this
case the Wino-like (triplet EWIMP) or the Higgsino-like (doublet
EWIMP) dark matter may be realized. The lighter shaded areas (as
bright as the shaded areas in Fig.~\ref{massdiff1}) are the
Higgsino-like region ($|Z_{13}|^2 + |Z_{14}|^2 > 0.9$), and the darker
shaded areas are the Wino-like one ($|Z_{12}|^2 > 0.9$). The Wino-like
neutralino is highly degenerate with the lightest chargino in mass as
expected.

When the triplet EWIMP mass is around $1.7$ TeV, the thermal relic
density of the dark matter is consistent with the WMAP data. In the
doublet EWIMP case, the mass around $1$ TeV explains the WMAP data
\cite{Profumo:2004at}. However, note that the dark matter in the universe
may be produced thermally \cite{Kolb:1990vq}, or non-thermally
\cite{Enqvist:1998en}. Therefore, we do not assume any scenarios for the
dark matter relic density in this paper. Instead, we assume the dark
matter forms the dark halo in our galaxy with the appropriate mass
density. 

\vspace{0.5cm}
\underline{2. Threshold singularity}
\vspace{0.5cm}

\begin{figure}[t]
 \begin{center}
  \includegraphics[height = 4cm,clip]{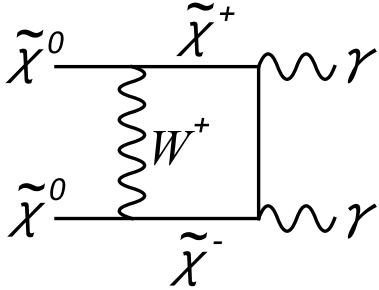}
 \end{center}
 \caption{\small
 Dominant diagram in the Wino- or Higgsino-like neutralino
 annihilation to two photons at one-loop level,
 when the neutralino is heavy compared to the weak gauge bosons.
 \label{fig3}
 }
\end{figure}

In the calculations of the EWIMP annihilation cross sections, a
threshold singularity appears due to the gauge interactions. For
investigating the singularity, let us consider the EWIMP annihilation 
cross section to two photons
$\sigma(\tilde{\chi}^0\tilde{\chi}^0\rightarrow \gamma\gamma)$ as an
example. The signal line gamma ray flux from the galactic center is
proportional to $\sigma(\tilde{\chi}^0\tilde{\chi}^0\rightarrow 
\gamma\gamma)$. This process is a radiative one, and the full one-loop
non-relativistic cross section in the MSSM context has already been
calculated in Ref.~\cite{Bergstrom:1997fh}. It is found that the cross
section is suppressed only by the $W$ boson mass, not by the
neutralino mass as
\begin{eqnarray}
 \sigma v
 \sim
 \frac{\alpha^2\alpha_2^2}{m_W^2}~,
 \label{1-loop result}
\end{eqnarray}
if the neutralino is heavy and almost Wino- or Higgsino-like. The
dominant diagram is shown in Fig.~\ref{fig3}.

\begin{figure}[t]
 \begin{center}
  \includegraphics[height = 3.5cm,clip]{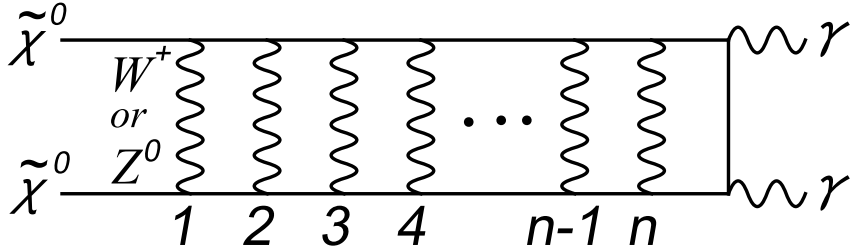}
 \end{center}
 \caption{\small
 Dominant diagram in the Wino- or Higgsino-like neutralino
 annihilation at ${\cal O}(\alpha\alpha_2^n)$, in which $n$ weak gauge bosons
 are exchanged. 
 \label{fig4}
 }
\end{figure}

On the other hand, the cross section must be bounded from above by
the unitarity limit,
\begin{eqnarray}
 \sigma v < \frac{4\pi}{v m^2}~.
 \label{unitarity bound}
\end{eqnarray}
Thus, the one-loop cross section exceeds the bound for the extremely heavy 
neutralino. It means that the higher-order corrections should be
included. The dominant higher-order contribution comes from the ladder
diagrams. The $n$-th order ($\alpha_2^n$) ladder diagram, in
which $n$ weak gauge bosons are exchanged, is depicted in
Fig.~\ref{fig4}. The corresponding amplitude ${\cal A}_n$ of the diagram
is roughly given by
\begin{eqnarray}
 {\cal A}_n
 \simeq
 \alpha
 \left(
  \frac{\alpha_2m}{m_W}
 \right)^n~.
\end{eqnarray}
When the neutralino mass  $m$ is large enough, the diagrams are enhanced
by a factor of $\alpha_2 m/m_W$ for each weak gauge boson exchange. The
higher-order loop diagrams become more and more important when $\alpha_2
m \gsim m_W$.

Enhancement of ladder diagrams in non-relativistic limits is related
to a threshold singularity. Recall that a threshold singularity
appears in the non-relativistic $\mu^+\mu^-$ pair annihilation cross
section. When the relative velocity $v$ of the muon pair is smaller
than $\alpha$, the amplitude of the $n$-order ladder diagram, in which
$n$ photons are exchanged between the muon pair, is proportional to
$\alpha (\alpha/v)^n$, and the perturbative expansion by $\alpha$
breaks down. The internal muons are close to non-relativistic on-shell
states. The muon and photon propagaters are proportional to $1/v^2$
and each loop integration gives $\alpha v^5$. Thus, the diagrams
are enhanced by $\alpha/v$ for each photon exchange.  This is because
the kinetic energy of muon pair, $m_\mu v^2/4$, is smaller than the
Coulomb potential energy, $
\alpha^2 m_\mu$, and the wave function of the incident particles is
deformed from plane waves. We need to systematically resum the ladder
diagrams or to use the wave function under the Coulomb potential in
order to get the precise annihilation cross section.

In the non-relativistic EWIMP pair annihilation, the sub-diagram
corresponding  to the process $\tilde{\chi}^0\tilde{\chi}^0\rightarrow
\tilde{\chi}^+\tilde{\chi}^-$ in each ladder diagram is very close to 
the threshold when the mass difference $\delta m$ is negligible.  In
this case, the spatial momentums for EWIMPs and the SU(2)$_L$
partners in the internal lines are regularized by the weak gauge boson
masses. Their propagaters and the weak gauge boson ones behave as $m/m_W^2$
and $1/m_W^2$, respectively, and the loop integration gives $\alpha_2
m_W^5/m$. Thus, the diagrams are enhanced by $\alpha_2 m/m_W$ for each
weak gauge boson exchange, as shown above.  This implies that when $\alpha_2
m\gsim m_W$, the weak interaction becomes a long-range force and the
wave function is significantly modified inside the Yukawa potentials
induced by the weak gauge boson exchanges in the non-relativistic
limit. In the following sections, we will introduce a systematic
method to derive the annihilation cross sections in the threshold
singularity region by using the non-relativistic action.

The elastic scattering cross section of dark matter with nucleon
is important for the direct search for dark matter
\cite{Griest:1988ma}. If the dark matter is an EWIMP and much heavier
than the weak gauge boson masses, the one-loop correction to the cross
section is suppressed only by the weak gauge boson masses and it may be
dominant over the tree-level contribution \cite{Hisano:2004pv}. However,
the perturbative expansion is still reliable, unlike the case of the
annihilation cross sections.  This is because the reduced mass in the
EWIMP and nucleon two-body system is not heavy enough for 
non-perturbative corrections to be sizable.


\vspace{1.0cm}
\lromn 3 \hspace{0.2cm} {\bf Two-body State Effective Action}
\vspace{0.5cm}

In this section we derive the effective actions for the
non-relativistic two-body states including pairs of EWIMPs and the
SU(2)$_L$ partners. The action is derived by following steps. (i) We
integrate out all fields except EWIMPs and the SU(2)$_L$ partners
such as the lightest neutralino and chargino in the MSSM action. (ii)
The non-relativistic action ${\cal S}_{\rm NR}$ is obtained by
integrating out large momentum modes of EWIMPs and the SU(2)$_L$
partners. The action includes the effect of the EWIMP annihilation
as the absorptive parts. (iii) The action ${\cal S}_{\rm NR}$ is expanded
by the velocities of EWIMPs. (iv) At last, we introduce auxiliary fields
to the action ${\cal S}_{\rm NR}$, which represent two-body states of
EWIMPs and the SU(2)$_L$ partners. The two-body state effective
action ${\cal S}^{(II)}$ is obtained by integrating out all fields
except those auxiliary fields in the non-relativistic action ${\cal
S}_{\rm NR}$.

In the following, we derive the two-body state effective action for
the triplet EWIMP (Wino-like neutralino). For the doublet
EWIMP (Higgsino-like neutralino), only the final result is presented,
because the derivation is essentially the same as that of the triplet
one.

\vspace{0.5cm}
\underline{1. Integrating out all fields except $\tilde{\chi}^0$ and
$\tilde{\chi}^-$}
\vspace{0.5cm}

Relevant interactions to the annihilation cross section for the
triplet EWIMP are the gauge interaction in Eq.~(\ref{gauge interaction
of Wino}). In the MSSM, the lightest neutralino and chargino have
other interactions with sfermions and Higgs bosons. However, the
contributions to the annihilation cross sections are suppressed by the
sfermion masses or the gaugino-Higgsino mixing. We assume that these
contributions are small, and neglect them.

After integrating out the gauge bosons, the effective action for the triplet
EWIMP and its charged SU(2)$_L$ partner becomes
\begin{eqnarray}
 {\cal S}_{\rm eff}
 &=&
 \int d^4x
 \left[
  \frac{1}{2}
  \overline{\tilde{\chi}^0}\left(i\SlashIt{\partial} - m\right)\tilde{\chi}^0
  +
  \overline{\tilde{\chi}^-}\left(i\SlashIt{\partial} - m_c\right)\tilde{\chi}^-
 \right]
 +
 {\cal S}_{\rm int}[\tilde{\chi}^0,\tilde{\chi}^-]~,
 \label{WSeff}
\end{eqnarray}
\begin{eqnarray}
 {\cal S}_{\rm int}
 &=&
 2i\pi\alpha\int d^4x_1 d^4x_2
 \left[~
  \frac{2}{s_W^2}
  j_W^{\mu\dagger}(x_1)
  ~G^{(W)}_{\mu\nu}(x_1 - x_2)~
  j_W^\nu(x_2)
 \right.
 \nonumber \\
 \nonumber \\
 &~&\qquad~~~
 \left.
  +~
  j_{\chi^-}^\mu(x_1)
  \left\{
   G^{(\gamma)}_{\mu\nu}(x_1 - x_2)
   +
   \frac{c_W^2}{s_W^2}G^{(Z)}_{\mu\nu}(x_1 - x_2)
  \right\}
  j_{\chi^-}^\mu(x_2)
 \right]~,
 \label{WSeff int}
\end{eqnarray}
where the parameter $m_c$ is the mass of the charged SU(2)$_L$
partner. The functions $G^{(W)}_{\mu\nu}(x)$, $G^{(Z)}_{\mu\nu}(x)$ and
$G^{(\gamma)}_{\mu\nu}(x)$ are the Feynman propagaters of the $W$, $Z$
bosons and photon, respectively. The currents $j_W^{\mu}(x)$ and
$j_{\chi^-}^\mu(x)$ are defined as
\begin{eqnarray}
 j_W^{\mu}(x)
 =
 \overline{\tilde{\chi}^0}(x)\gamma^\mu\tilde{\chi}^-(x)~,
 \qquad
 j_{\chi^-}^\mu(x)
 =
 \overline{\tilde{\chi}^-}(x)\gamma^\mu\tilde{\chi}^-(x)~.
\end{eqnarray}

Here we include the effects of the non-vanishing mass difference between
the EWIMP and its SU(2)$_L$ partner, $\delta m(\equiv m_c-m)$, which
comes from the electroweak symmetry breaking. The non-vanishing
$\delta m$ gives sizable effects on the annihilation cross sections
when $\delta m$ is not negligible compared with $\alpha_2 m_W$ as will
be shown in Sec.~\lromn 5.  On the other hand, the
SU(2)$_L\otimes$U(1)$_Y$ breaking in the gauge interactions gives at
most corrections up to $O(m_W^2/\Lambda^2)$ to the cross sections,
and they can be ignored as far as $m_W\ll\Lambda$.

\vspace{0.5cm}
\underline{2. Integrating out large momentum modes of 
$\tilde{\chi}^0$ and $\tilde{\chi}^-$}
\vspace{0.5cm}

We now derive the action which describes the non-relativistic motion of
the EWIMP and its SU(2)$_L$ partner. Namely we integrate out the large
momentum modes of these particles in ${\cal S}_{\rm eff}$. We divide the
fields $\tilde{\chi}^0$ and $\tilde{\chi}^-$ into two parts, the
non-relativistic part and the other, 
\begin{eqnarray}
 \tilde{\chi}^0(x)
 &=&
 \tilde{\chi}^0_{\rm NR}(x)
 +
 \delta\tilde{\chi}^0(x)~,
 \nonumber \\
 \nonumber \\
 \tilde{\chi}^0_{\rm NR}(x)
 &=&
 \int_{[{\rm NR}]} \frac{d^4p}{(2\pi)^4} \phi^0(p)e^{-ipx}~,
 \qquad
 \delta\tilde{\chi}^0(x)
 =
 \int_{[{\rm \overline{\rm NR}}]}
 \frac{d^4p}{(2\pi)^4} \phi^0(p)e^{-ipx}~,
 \label{W-division}
\end{eqnarray}
where the $\phi^0(p)$ is the Fourier coefficient of the EWIMP field. The
region of the integration $[{\rm NR}]$ is defined roughly by $[{\rm NR}]
= \{(p^0, \vec{p})~|~p^0 = \pm m + {\cal O}(m|\vec{v}|^2),~\vec{p} =
{\cal O}(m\vec{v}),~|\vec{v}|\ll 1 \}$, and $[\overline{\rm NR}]$ means
the complementary set of $[{\rm NR}]$. The SU(2)$_L$ partner field 
$\tilde{\chi}^-$ is also divided into $\tilde{\chi}^-_{\rm NR}$ and
$\delta \tilde{\chi}^-$ in the same way. After integrating out large
momentum modes $\delta\tilde{\chi}^0$ and $\delta \tilde{\chi}^-$ in the
action ${\cal S}_{\rm eff}$, the non-relativistic effective action
${\cal S}_{\rm NR}$ is obtained as
\begin{eqnarray}
 {\cal S}_{\rm NR}
 &=&
 \int d^4x
 \left[
  \frac{1}{2}
  \overline{\tilde{\chi}^0}_{\rm NR}
  \left(i\SlashIt{\partial} - m\right)
  \tilde{\chi}^0_{\rm NR}
  +
  \overline{\tilde{\chi}^-}_{\rm NR}
  \left(i\SlashIt{\partial} - m_c\right)
  \tilde{\chi}^-_{\rm NR}
 \right]
 \nonumber \\
 \nonumber \\
 &~&
 +
 {\cal S}_{\rm int}[\tilde{\chi}^0_{\rm NR},\tilde{\chi}^-_{\rm NR}]
 +
 \delta{\cal S}[\tilde{\chi}^0_{\rm NR},\tilde{\chi}^-_{\rm NR}]~.
 \label{WNR}
\end{eqnarray}
All effective interactions induced from the integration by
$\delta\tilde{\chi}^0$ and $\delta\tilde{\chi}^-$ are included in
$\delta{\cal S}$. Though they are suppressed by the EWIMP mass in
comparison with interactions in ${\cal S}_{\rm int}$, they give 
leading contributions to the imaginary part (absorptive part) of the
non-relativistic action. The action is further simplified in the next
step.

\vspace{0.5cm}
\underline{3. Non-relativistic expansion of the action ${\cal S}_{\rm NR}$} 
\vspace{0.5cm}

Here, we expand the action ${\cal S}_{\rm NR}$ by the velocity
of the EWIMP. For the expansion, it is convenient
to use two-components spinor fields $\zeta$, $\eta$ and $\xi$ instead
of $\tilde{\chi}^0_{\rm NR}$ and $\tilde{\chi}^-_{\rm NR}$. These
spinor fields are defined by
\begin{eqnarray}
 \tilde{\chi}^0_{\rm NR}
 =
 \left(
  \begin{array}{r}
   e^{-imt}\zeta
   +
   ie^{ imt}\ds\frac{\vec{\nabla}\cdot\vec{\sigma}}{2m} \zeta^c
   \\
   \\
   e^{ imt} \zeta^c
   -
   ie^{-imt}\ds\frac{\vec{\nabla}\cdot\vec{\sigma}}{2m}\zeta
  \end{array}
 \right)~,
 \qquad
 \tilde{\chi}^-_{\rm NR}
 =
 \left(
  \begin{array}{r}
   e^{-imt}\eta
   +
   ie^{ imt}\ds\frac{\vec{\nabla}\cdot\vec{\sigma}}{2m} \xi
   \\
   \\
   e^{ imt} \xi
   -
   ie^{-imt}\ds\frac{\vec{\nabla}\cdot\vec{\sigma}}{2m}\eta
  \end{array}
 \right)~.
\end{eqnarray}
The spinor $\zeta^c$ is the charge conjugation of $\zeta$, $\zeta^c =
-i\sigma^2\zeta^*$, where $\sigma_2$ is the Pauli matrix. Spinors $\zeta$
and $\eta$ annihilate one $\tilde{\chi}^0$ and one $\tilde{\chi}^-$,
respectively, while $\xi$ creates one $\tilde{\chi}^+$.

The non-relativistic action ${\cal S}_{\rm NR}$ is systematically
expanded by the velocity of the two-components spinor fields. The
kinetic terms in Eq.~(\ref{WNR}) become
\begin{eqnarray}
 \left.S_{\rm NR}\right|_{\rm kinetic~terms}
 &=&
 \int d^4x
 \left[~
  \zeta^\dagger
  \left(
   i\partial_t + \frac{\nabla^2}{2m}
  \right)\zeta
  +
  \eta^\dagger
  \left(
   i\partial_t - \delta m + \frac{\nabla^2}{2m}
  \right)\eta
 \right.
 \nonumber \\
 \nonumber \\
 &~&\qquad\qquad\qquad\qquad\qquad~~~
 \left.
  +~
  \xi^\dagger
  \left(
   i\partial_t + \delta m - \frac{\nabla^2}{2m}
  \right)\xi~
 \right]~.
 \label{KT}
\end{eqnarray}
The interactions in ${\cal S}_{\rm int}$ of Eq.~(\ref{WNR}) are reduced as
\begin{eqnarray}
 {\cal S}_{\rm int}
 &=&
 \int d^4xd^3y
 \left[~~~~~
  \frac{\alpha}{2|\vec{x} - \vec{y}|}
  \left(
   1 + \frac{c_W^2}{s_W^2}e^{-m_Z|\vec{x} - \vec{y}|}
  \right)
  \eta^\dagger(x)\xi(\vec{y},x^0)
  \times
  \xi^\dagger(\vec{y},x^0)\eta(x)
 \right.
 \nonumber \\
 \nonumber \\
 &~&\qquad\qquad~~~
 \left.
  +\frac{\alpha_2 e^{-m_W|\vec{x} - \vec{y}|}}{2|\vec{x} - \vec{y}|}
  \left\{
   \zeta^\dagger(x)\zeta^c(\vec{y},x^0)
   \times
   \xi^\dagger(\vec{y},x^0)\eta(x)
   +
   {\rm h.c.}
  \right\}~~
 \right]~,
 \label{POT}
\end{eqnarray}
where we keep terms which dictate the transitions between states
with both spin and isospin singlet. The EWIMP $S$-wave state is
spin-singlet due to the Majorana nature, and only those terms are kept
to calculate the EWIMP annihilation cross sections at the
non-relativistic limit. The first term in the parenthesis describes the
Coulomb force and the force by one $Z$-boson exchange between
$\tilde{\chi}^+$ and $\tilde{\chi}^-$. The second term is for the
transition between $\tilde{\chi}^0\tilde{\chi}^0$ and
$\tilde{\chi}^+\tilde{\chi}^-$ by one $W$ boson exchange. 

The imaginary part (absorptive part) of the action ${\cal S}_{\rm NR}$
is in $\delta{\cal S}$ in Eq.~(\ref{WNR}). It comes from following box
diagrams; (a) transitions from $\tilde{\chi}^+\tilde{\chi}^-$ to
$\tilde{\chi}^+\tilde{\chi}^-$ with intermediate $W^+W^-$, $Z^0Z^0$,
$\gamma Z$ and $\gamma\gamma$ states, (b) a transition from
$\tilde{\chi}^0\tilde{\chi}^0$ to $\tilde{\chi}^0\tilde{\chi}^0$ with
an intermediate $W^+W^-$ state and (c) a transition from
$\tilde{\chi}^0\tilde{\chi}^0$ to $\tilde{\chi}^+\tilde{\chi}^-$ with
an intermediate $W^+W^-$ state. These effective interactions are
simplified in the non-relativistic expansion as
\begin{eqnarray}
 \delta{\cal S}
 &=&
 \frac{i\pi\alpha_2^2}{m^2}\int d^4x
 \left[~~
  \left(
   \frac{1}{2}
   +
   c_W^4
   +
   2s_W^2c_W^2
   +
   s_W^4
  \right)
  \eta^\dagger\xi\cdot\xi^\dagger\eta
 \right.
 \nonumber \\
 \nonumber \\
 &~&\qquad\qquad\qquad
 \left.
  +
  2
  \zeta^\dagger\zeta^c\cdot\zeta^{c\dagger}\zeta
  +
  \frac{1}{2}
  \left(
   \zeta^\dagger\zeta^c\cdot\xi^\dagger\eta
   +
   {\rm h.c.}
  \right)~~
 \right]~.
 \label{ANN}
\end{eqnarray}
Here, we assume that the EWIMP mass $m$ is much heavier than the weak
gauge boson masses, $m\gg m_W,m_Z$. The first term of the right-hand
side in Eq.~(\ref{ANN}) corresponds to the box diagrams (a). Each term
in the parenthesis comes from the diagrams with intermediate $W^+W^-$,
$Z^0Z^0$, $\gamma Z$ and $\gamma\gamma$ states, respectively. The second and
third terms correspond to the diagrams of (b) and (c), respectively. 

\vspace{0.5cm}
\underline{4. Two-body state effective action} 
\vspace{0.5cm}

The non-relativistic action for the triplet EWIMP is now given by
Eqs.~(\ref{KT} - \ref{ANN}). We now introduce auxiliary fields
$\sigma_N$ and $\sigma_C$, which describe the two-body states
$\tilde{\chi}^0\tilde{\chi}^0$ and $\tilde{\chi}^+\tilde{\chi}^-$ with
both spin and isospin singlet, respectively. We thus insert identities 
\begin{eqnarray}
 1
 &=&
 \int {\cal D}\sigma_N {\cal D}s^\dagger_N
 \exp
 \left[
  \frac{i}{2}\int d^4xd^3y~
  \sigma_N(x,\vec{y})
  \left\{
   s_N^\dagger(\vec{y},x) - \frac{1}{2}\zeta^\dagger(x)\zeta^c(\vec{y},x^0)
  \right\}
 \right]~,
 \nonumber \\
 \nonumber \\
 1
 &=&
 \int {\cal D}\sigma_C {\cal D}s^\dagger_C
 \exp
 \left[
  \frac{i}{2}\int d^4xd^3y~
  \sigma_C(x,\vec{y})
  \left\{
   s_C^\dagger(\vec{y},x) - \eta^\dagger(x)\xi(\vec{y},x^0)
  \right\}
 \right]~,
\end{eqnarray}
and their conjugates into the partition function described by the
non-relativistic action. After integrating out $\eta$, $\xi$, $\zeta$,
$s_N$, $s_C$ and their conjugates, the two-body state effective action
${\cal S}^{(II)}$ is obtained as  
\begin{eqnarray}
 {\cal S}^{(II)}
 =
 \int d^4xd^3r~
 \Phi^\dagger(x,\vec{r})
 \left\{
  \left(
   i\partial_{x^0} + \frac{\nabla_x^2}{4m} + \frac{\nabla_r^2}{m}
  \right)
  -
  {\bf V}(\vec{r})
  +
  2i{\bf \Gamma}\delta(\vec{r})
 \right\}\Phi(x,\vec{r})~,
 \label{W2bodyS}
\end{eqnarray}
where the argument $x$ denotes the center of mass coordinate in the
two-body system and $\vec{r}$ is the relative coordinate. The
two-components two-body state field $\Phi(x,\vec{r})$ is defined by 
\begin{eqnarray}
 \Phi(x,\vec{r})
 =
 \left(
  \begin{array}{c}
   \phi_C(x,\vec{r})
   \\
   \phi_N(x,\vec{r})
  \end{array}
 \right)
 =
 \left(- {\bf V}(r) + 2i{\bf \Gamma}\delta(\vec{r})\right)^{-1}
 \left(
  \begin{array}{c}
   \sigma_C(x,\vec{r})
   \\
   \sigma_N(x,\vec{r})
  \end{array}
 \right)~.
\end{eqnarray}
The components, $\phi_N$ and $\phi_C$, describe pairs of the EWIMPs and
the SU(2)$_L$ partners, respectively. These can be also written as 
\begin{eqnarray}
 \phi_C(x,\vec{r})
 &=&
 \frac{1}{\sqrt{2}}
 \xi^\dagger(\vec{x} - \vec{r}/2,x^0)~
 \eta(\vec{x} + \vec{r}/2,x^0)~,
 \label{normalization for C}
 \\
 \nonumber \\
 \phi_N(x,\vec{r})
 &=&
 \frac{1}{2}
 \zeta^{c\dagger}(\vec{x} - \vec{r}/2,x^0)~
 \zeta(\vec{x} + \vec{r}/2,x^0)~,
 \label{normalization for N}
\end{eqnarray}
by using $\zeta$, $\eta$ and $\xi$ fields. The difference between the
normalizations of $\phi_C$ in Eq.~(\ref{normalization for C}) and
$\phi_N$ in Eq.~(\ref{normalization for N}) comes from the fact that
$\phi_N$ describes a pair of the identical Majorana fermions. 

The electroweak potential ${\bf V}(r)$ in Eq.~(\ref{W2bodyS}) is given
by 
\begin{eqnarray}
 {\bf V}(r)
 =
 \begin{pmatrix}
  2\delta m - \ds\frac{\alpha}{r} - \alpha_2c_W^2\ds\frac{e^{-m_Zr}}{r}
  &
  -\ds\sqrt{2}\alpha_2\ds\frac{e^{-m_Wr}}{r}
  \\
  -\ds\sqrt{2}\alpha_2\ds\frac{e^{-m_Wr}}{r}
  &
  0
 \end{pmatrix}~.
 \label{WV}
\end{eqnarray}
The absorptive (imaginary) part ${\bf \Gamma}$ is decomposed as ${\bf \Gamma}
= {\bf \Gamma}_{W^+W^-} + {\bf \Gamma}_{Z^0Z^0} + {\bf \Gamma}_{\gamma
Z^0} + {\bf \Gamma}_{\gamma\gamma}$ where each component is
\begin{eqnarray}
 {\bf \Gamma}_{W^+W^-}
 &=&
 \frac{\pi\alpha_2^2}{4m^2}
 \begin{pmatrix}
  2 & \sqrt{2}
  \\
  \sqrt{2} & 4
 \end{pmatrix}~,
 \qquad
 {\bf \Gamma}_{Z^0Z^0}
 =
 \frac{\pi\alpha_2^2}{m^2}
 \begin{pmatrix}
  c_W^4 & 0
  \\
  0 & 0
 \end{pmatrix}~,
 \nonumber\\
 {\bf \Gamma}_{\gamma Z^0}
 &=&
 \frac{\pi\alpha\alpha_2}{m^2}
 \begin{pmatrix}
  2c_W^2 & 0
  \\
  0 & 0
 \end{pmatrix}~,
 \qquad
 {\bf \Gamma}_{\gamma\gamma}
 =
 \frac{\pi\alpha^2}{m^2}
 \begin{pmatrix}
  1 & 0
  \\
  0 & 0
 \end{pmatrix}~.
 \label{WG}
\end{eqnarray}
The two-body state effective action Eq.~(\ref{W2bodyS}) is the final
result of this section. 
 
In the case of the doublet EWIMP, the derivation of the effective action
is parallel to that of the triplet one. The EWIMP is accompanied with
the neutral SU(2)$_L$ partner in addition to the charged one. The
two-body effective action becomes $3\times 3$ matrix form, and the
two-body state field $\Phi(x,\vec{r})$ has three components, 
\begin{eqnarray}
 \Phi(x,\vec{r})
 =
 \left(
  \begin{array}{c}
   \phi_C(x,\vec{r})
   \\
   \phi_N(x,\vec{r})
   \\
   \phi_{N_2}(x,\vec{r})
  \end{array}
 \right)~,
\end{eqnarray}
where $\phi_N$, $\phi_C$ and $\phi_{N_2}$ describe pairs of the EWIMPs,
the charged and the neutral partners, respectively. The electroweak
potential is 
\begin{eqnarray}
 {\bf V}(r)
 =
 \begin{pmatrix}
  2\delta m
  -\ds\frac{\alpha}{r}
  -\ds\frac{\alpha_2(1 - 2c_W^2)^2}{4c_W^2}\frac{e^{-m_Zr}}{r}
  &
  -\ds\frac{\sqrt{2}\alpha_2e^{-m_Wr}}{4r}
  &
  -\ds\frac{\sqrt{2}\alpha_2e^{-m_Wr}}{4r}
  \\
  -\ds\frac{\sqrt{2}\alpha_2e^{-m_Wr}}{4r}
  &
  0
  &
  -\ds\frac{\alpha_2e^{-m_Zr}}{4c_W^2r}
  \\
  -\ds\frac{\sqrt{2}\alpha_2e^{-m_Wr}}{4r}
  &
  -\ds\frac{\alpha_2e^{-m_Zr}}{4c_W^2r}
  &
  2\delta m_N
 \end{pmatrix}~,
 \label{HV}
\end{eqnarray}
where $\delta m_N$ is the mass difference between the EWIMP and its
neutral SU(2)$_L$ partner. The absorptive part ${\bf \Gamma}$ in this case is
given by ${\bf \Gamma} = {\bf \Gamma}_{W^+W^-} + {\bf \Gamma}_{Z^0Z^0} +
{\bf \Gamma}_{\gamma Z^0} + {\bf \Gamma}_{\gamma\gamma}$, where 
\begin{eqnarray}
 {\bf \Gamma}_{W^+W^-}
 &=&
 \frac{\pi\alpha_2^2}{64m^2}
 \begin{pmatrix}
  8 & \sqrt{2} & \sqrt{2}
  \\
  \sqrt{2} & 4 & 4
  \\
  \sqrt{2} & 4 & 4
 \end{pmatrix}~,
 \nonumber \\
 \qquad
 {\bf \Gamma}_{Z^0Z^0}
 &=&
 \frac{\pi\alpha_2^2}{64c_W^4m^2}
 \begin{pmatrix}
  4(1 - 2s_W^2)^4 & \sqrt{2}(1 - 2s_W^2)^2 & \sqrt{2}(1 - 2s_W^2)^2
  \\
  \sqrt{2}(1 - 2s_W^2)^2 & 2 & 2
  \\
  \sqrt{2}(1 - 2s_W^2)^2 & 2 & 2
 \end{pmatrix}~,
 \nonumber\\
 {\bf \Gamma}_{\gamma Z^0}
 &=&
 \frac{\pi\alpha\alpha_2}{2c_W^2m^2}
 \begin{pmatrix}
  (1 - 2s_W^2)^2 & 0 & 0
  \\
  0 & 0 & 0
  \\
  0 & 0 & 0
 \end{pmatrix}~,
 \qquad
 {\bf \Gamma}_{\gamma\gamma}
 =
 \frac{\pi\alpha^2}{m^2}
 \begin{pmatrix}
  1 & 0 & 0
  \\
  0 & 0 & 0
  \\
  0 & 0 & 0
 \end{pmatrix}~.
 \label{HG}
\end{eqnarray}


\vspace{1.0cm}
\lromn 4 \hspace{0.2cm} {\bf Optical Theorem and Annihilation Cross Sections}
\vspace{0.5cm}

We now derive the EWIMP pair annihilation cross sections  by using
the two-body state effective actions in Eq.~(\ref{W2bodyS}). Using the
formula, we also show that the one-loop cross 
section of $\tilde{\chi}^0\tilde{\chi}^0\rightarrow 2\gamma$ is
reproduced in the perturbative expression while
the cross section in a limit of $m\rightarrow\infty$ satisfies the 
unitarity bound.

\vspace{0.5cm}
\underline{1. Annihilation cross section formula}
\vspace{0.5cm}

Due to the optical theorem, the total EWIMP pair annihilation cross section
is written by the imaginary part of the amplitude as
\begin{eqnarray}
 \sqrt{s^2 - 4m^2s}~\sigma
 =
 \Im\left[{\cal M}_{ii}\right]~,
 \label{Optical Theorem}
\end{eqnarray}
where $\sqrt{s}$ is the center of mass energy, and ${\cal M}_{ii}$ is
the invariant amplitude of the process,
$\tilde{\chi}^0\tilde{\chi}^0\rightarrow\tilde{\chi}^0\tilde{\chi}^0$.
When the  incident EWIMPs are highly non-relativistic, the
relevant initial state for the pair annihilation is only the
spin-singlet $S$-wave state as mentioned before. We thus project the
above 
equation (\ref{Optical Theorem}) to the $^1S_0$ state by using the
projection operator $\int d^3Pdk |\vec{P},k,~^1S_0; N\rangle\langle
\vec{P},k,~^1S_0; N|$, 
\begin{eqnarray}
 |\vec{P},k,~^1S_0; N\rangle
 &=&
 \frac{k}{4\sqrt{\pi}}
 \int d\Omega_k
 \left[~~~
  a^\dagger_+(\vec{P}/2 + \vec{k})~
  a^\dagger_-(\vec{P}/2 - \vec{k})
 \right.
 \nonumber \\
 &&\qquad\qquad\qquad
 \left.
  -
  a^\dagger_-(\vec{P}/2 + \vec{k})~
  a^\dagger_+(\vec{P}/2 - \vec{k})
 \right]
 |0\rangle~,
 \label{P for N}
\end{eqnarray}
where $\vec{P}$ is the total momentum and $\vec{k}$ is the relative
momentum in the EWIMP two-body system. The variable $k$ is $k =
|\vec{k}|$. The state vector $|\vec{P},k,^1S_0\rangle$ is normalized as
$\langle \vec{P}',k',~^1S_0; N|\vec{P},k,~^1S_0; N\rangle =
\delta(\vec{P} - \vec{P}') \delta(k - k')$. The operator
$a^\dagger_{\pm}(\vec{k})$ creates one $\tilde{\chi}^0$  with the momentum 
$\vec{k}$ and the spin ${\pm}1/2$, and satisfies the anti-commutation
relation $\{a_{s}(\vec{p}),a^\dagger_{s'}(\vec{p}')\} = \delta(\vec{p} -
\vec{p}')\delta_{ss'}$.

If the $\tilde{\chi}^+\tilde{\chi}^-$ annihilation cross section is
considered in a spin singlet and $S$-wave system, the projection
operator $\int d^3Pdk |\vec{P},k,~^1S_0;
C\rangle\langle\vec{P},k,^1S_0; C|$ is used instead of one in
Eq.~(\ref{P for N}), where
\begin{eqnarray}
 |\vec{P},k,~^1S_0; C\rangle
 &=&
 \frac{k}{2\sqrt{2\pi}}
 \int d\Omega_k
 \left[~~~
  b^\dagger_+(\vec{P}/2 + \vec{k})~
  d^\dagger_-(\vec{P}/2 - \vec{k})
 \right.
 \nonumber \\
 &&\qquad\qquad\qquad~
 \left.
  -
  b^\dagger_-(\vec{P}/2 + \vec{k})~
  d^\dagger_+(\vec{P}/2 - \vec{k})
 \right]
 |0\rangle~.
 \label{P for C}
\end{eqnarray}
The operators $b^\dagger_{\pm}(\vec{k})$ and
$d^\dagger_{\pm}(\vec{k})$ create one $\tilde{\chi}^-$ and one
$\tilde{\chi}^+$ with the momentum $\vec{k}$ and the spin
${\pm}1/2$, respectively. They satisfy the anti-commutation
relations $\{b_{s}(\vec{p}),b^\dagger_{s'}(\vec{p}')\} =
\{d_{s}(\vec{p}),d^\dagger_{s'}(\vec{p}')\} = \delta(\vec{p} -
\vec{p}')\delta_{ss'}$.

After the projection, the $S$-wave cross sections $\sigma^{(S)}_i$ $(i =
1,2)$ become
\begin{eqnarray}
 \sigma^{(S)}_i
 &=&
 c_i\frac{32\pi^5}{m^2v^3}~
 \Im\left[{\cal M}^{(S)}_i(v)\right]~,
\end{eqnarray}
where $v$ is the relative velocity between incident particles. The cross 
section $\sigma^{(S)}_i$ is for $\tilde{\chi}^0\tilde{\chi}^0$
annihilation ($i = 2$) and $\tilde{\chi}^+\tilde{\chi}^-$ annihilation
($i = 1$). The coefficient $c_i$ is given by $c_1 = 1$ and $c_2 =
2$. ${\cal M}^{(S)}_i(v)$ is
the invariant amplitude from the $^1S_0$ state to the $^1S_0$ state. We
use the non-relativistic approximation as $s\simeq 4m^2 + m^2v^2$
for deriving the cross sections.

The invariant amplitude ${\cal M}^{(S)}_i(v)$ is obtained from the
$S$-wave Green function of the two-body state field $\Phi$. From the 
two-body state effective action in Eq.~(\ref{W2bodyS}), the
Schwinger-Dyson equation (equation of motion 
for the Green function) is derived as 
\begin{eqnarray}
 &~&
 \left(
  i\partial_{x^0} + \frac{\nabla^2_x}{4m} 
  +
  \frac{\nabla^2_{r}}{m}
  -
  {\bf V}(r)
  +
  i{\bf \Gamma}\frac{\delta(r)}{2\pi r^2}
 \right)
 \vev{T\Phi(x,\vec{r})\Phi^\dagger(y,\vec{r}')}
 \nonumber \\
 &~&
 \qquad\qquad\qquad\qquad\qquad\qquad
 \qquad\qquad\qquad
 =~
 i\delta^{(4)}(x - y)\delta^{(3)}(\vec{r} - \vec{r}')~.
\end{eqnarray}
The potential ${\bf V}(r)$ and the absorptive part ${\bf \Gamma}$ are
defined in Eqs. (\ref{WV}) and (\ref{WG}) for the triplet case, and in
Eqs. (\ref{HV}) and (\ref{HG}) for the doublet case. Since the potential
depends on only $r$, the Green function can be expanded by
the Legendre Polynomials as
\begin{eqnarray}
 \vev{T\Phi(x,\vec{r})\Phi^\dagger(y,\vec{r}')}
 =
 \int\frac{d^4P}{(2\pi)^4}e^{-iP(x - y)}
 \sum_{l}\frac{2l + 1}{4\pi}P_l(\cos\gamma)(-i){\bf G}^{(E,l)}(r,r')~,
\end{eqnarray}
where $\gamma$ is the angle between $\vec{r}$ and $\vec{r}'$. The
variable $E$ in the superscript is the internal energy of the
two-body state ($E = P^0 - {\vec P}^2/4m$). The equation for the function
${\bf G}^{(E,l)}(r,r')$ is then given by 
\begin{eqnarray}
 \left(
  - E
  - \frac{1}{mr}\frac{d^2}{dr^2}r
  - \frac{l(l+1)}{mr^2} + {\bf V}(r)
  - i{\bf \Gamma}\frac{\delta(r)}{2\pi r^2}
 \right)
 {\bf G}^{(E,l)}(r,r')
 =
 \frac{\delta(r - r')}{r^2}~.
 \label{Schrodinger eq}
\end{eqnarray}

The invariant amplitude ${\cal M}^{(S)}_i(v)$ is written by the
$(i,i)$-component of the $S$-wave Green function ${\bf G}^{(E,0)}_{ii}$
as 
\begin{eqnarray}
 {\cal M}^{(S)}_i(v)
 =
 \frac{k^2}{4\pi^4}
 \lim_{E\rightarrow k^2/m}
 \left(E - \frac{k^2}{m}\right)^2
 \int_0^\infty r^2dr~{r'}^2dr'
 j_0(kr)j_0(kr')
 {\bf G}^{(E,0)}_{ii}(r,r')~,
\end{eqnarray}
where $k = mv/2$ and $j_0(x)$ is the zero-th order of the spherical Bessel
function. By using this expression, we obtain the formula for the
total $S$-wave annihilation cross section $\sigma^{(S)}_i$, 
\begin{eqnarray}
 \sigma^{(S)}_iv
 =
 c_i\frac{2\pi}{k^2}
 \lim_{E\rightarrow k^2/m}
 \left(E - \frac{k^2}{m}\right)^2
 \int_0^\infty rdr~r'dr'
 \sin(kr)\sin(kr')
 \Im\left[{\bf G}^{(E,0)}_{ii}(r,r')\right]~.
 \label{CS formula 1}
\end{eqnarray}
In general, the function ${\bf G}^{(E,0)}_{ii}(r,r')$ can not be solved 
analytically, therefore we need to solve the Schwinger-Dyson equation
(\ref{Schrodinger eq}) numerically to obtain the cross sections.

\vspace{0.5cm}
\underline{2. Solving the Schroedinger equation}
\vspace{0.5cm}

When a function, ${\bf g}(r,r') = rr'{\bf G}^{(E,0)}(r,r')$, is defined,
Eq.~(\ref{Schrodinger eq}) becomes the standard Schroedinger
equation in one dimension, 
\begin{eqnarray}
 -
 \frac{1}{m}\frac{d^2}{dr^2}{\bf g}(r,r')
 +
 \left(
  {\bf V}(r) - i{\bf \Gamma}\frac{\delta(r)}{2\pi r^2}
 \right){\bf g}(r,r')
 -
 E{\bf g}(r,r')
 =
 \delta(r - r')~.
 \label{1dimS}
\end{eqnarray}
In the following we expand the solution ${\bf g}(r,r')$ perturbatively by the
absorptive part ${\bf \Gamma}$.

At the leading order, the solution ${\bf
g}_0(r,r')$ satisfies the equation, 
\begin{eqnarray}
 -\frac{1}{m}\frac{d^2}{dr^2}{\bf g}_0(r,r')
 +{\bf V}(r){\bf g}_0(r,r')
 - E{\bf g}_0(r,r')
 =
 \delta(r - r')~.
 \label{eqg0}
\end{eqnarray}
The boundary conditions of the equation are determined by the following two
requirements; (i) The Green function ${\bf G}^{(E,0)}(r,r')$ must be
finite for any $r$ and $r'$. (ii) The Green function has only an
out-going wave at $|r - r'|
\rightarrow \infty$. Then the solution ${\bf g}_0(r,r')$ is obtained as
\begin{eqnarray}
 &~&
 {\bf g}_0(r,r')
 =
 m{\bf g}_>(r){\bf g}^T_<(r')\theta(r - r')
 +
 m{\bf g}_<(r){\bf g}^T_>(r')\theta(r' - r)~,
 \label{g0}
\end{eqnarray}
where ${\bf g}_{>(<)}(r)$ is the solution of the homogeneous equation,
\begin{eqnarray}
 -\frac{1}{m}\frac{d^2}{dr^2}{\bf g}_{>(<)}(r)
 +{\bf V}(r){\bf g}_{>(<)}(r)
 =
 E{\bf g}_{>(<)}(r)~.
 \label{Schrodinger}
\end{eqnarray}
Here, ${\bf g}_{>(<)}$ is given by a 2$\times$2 (3$\times$3) matrix in
the triplet (doublet) EWIMP case. The solution ${\bf g}_<(r)$ satisfies
the boundary conditions, (i) ${\bf g}_<(0) = {\bf 0}$ and (ii) ${\bf
g}'_<(0) = {\bf 1}$, while the conditions for ${\bf g}_>(r)$ are (i)
${\bf g}_>(0) = {\bf 1}$ and (ii) ${\bf g}_>(r)$  has only an out-going
wave at $r\rightarrow\infty$.

At the first order of ${\bf \Gamma}$, the solution of
Eq.~(\ref{1dimS}), ${\bf g}_1(r,r')$, is simply given by the
leading-order solution ${\bf g}_0(r,r')$ as
\begin{eqnarray}
 {\bf g}_1(r,r')
 =
 -\int dr''{\bf g}_0(r,r'')
 \left(
  -i{\bf \Gamma}\frac{\delta(r'')}{2\pi {r''}^2}
 \right)
 {\bf g}_0(r'',r')
 =
 \frac{im^2}{2\pi}{\bf g}_>(r)~{\bf \Gamma}~{\bf g}_>(r')~.
 \label{g1}
\end{eqnarray}

The $S$-wave cross sections in Eq.~(\ref{CS formula 1}) are proportional
to the imaginary part of the Green function ${\bf G}^{(E,0)}_{ii}$. The
imaginary part is related to not only the annihilation cross section but
also the elastic cross section of the process,
$\tilde{\chi}^0\tilde{\chi}^0 \rightarrow \tilde{\chi}^0\tilde{\chi}^0$
or $\tilde{\chi}^+\tilde{\chi}^- \rightarrow
\tilde{\chi}^+\tilde{\chi}^-$. After extracting the contribution of 
the annihilation processes
from $\Im\left[{\bf G}^{(E,0)}_{ii}\right]$, the total $S$-wave annihilation
cross section is obtained as
\begin{eqnarray}
 \sigma^{(S)}_iv
 &=&
 c_i\frac{m^2}{k^2}
 \lim_{E\rightarrow k^2/m}
 \left(E - \frac{k^2}{m}\right)^2
 \sum_{a,b}
 {\bf \Gamma}_{ab}{\cal A}_{ia}{\cal A}^*_{ib}~,
 \nonumber \\
 {\cal A}_{ia}
 &=&
 \int_0^\infty dr
 \sin(kr)\left[{\bf g}_>(r)\right]_{ia}~.
 \label{Cross2}
\end{eqnarray}

It is found that only the asymptotic behavior of the function ${\bf
g}_>(r)$ is relevant 
in the calculation of the cross sections. The function ${\bf g}_>(r)$
has only an out-going wave at $r\rightarrow\infty$ as stated above. When
the SU(2)$_L$ partners do not appear in the asymptotic state ($E <
2\delta m$), the function ${\bf g}_>(r)$ should behave as 
\begin{eqnarray}
 \left.
 \left[{\bf g}_>(r)\right]_{ij}
 \right|_{r\rightarrow\infty}
 &=&
 \delta_{i2}~d_{2j}(E)~e^{i\sqrt{mE}r}~.
\end{eqnarray}
Thus, the total EWIMP pair annihilation cross section  is given in a
simple form,
\begin{eqnarray}
 \sigma^{(S)}_2v
 &=&
 2\sum_{a,b}
 {\bf \Gamma}_{ab}~d_{2a}(mv^2/4)~d^*_{2b}(mv^2/4)~.
 \label{CS}
\end{eqnarray}

If the potential term in Eq.~(\ref{eqg0}) is neglected (in other word,
the long-distance effects are negligible), the coefficient $d_{2a}(E)$
is given by  $\delta_{2a}$ and the cross section is simply given by
$\sigma^{(S)}_2v = 2{\bf \Gamma}_{22}$, which is consistent with the
tree-level cross section as expected. Below we omit the subscript $2$
for the EWIMP annihilation cross section for simplicity.

\vspace{0.5cm}
\underline{3. One-loop result in
$\tilde{\chi}^0\tilde{\chi}^0\rightarrow  \gamma \gamma$ process}
\vspace{0.5cm}

For demonstration of the validity of Eq.~(\ref{CS}), we show that the
EWIMP annihilation cross section to two photons agrees with the one-loop
cross section when the potential term ${\bf V}$ is treated
perturbatively.

From Eq.~(\ref{CS}), the annihilation cross section to two photons is
given by 
\begin{eqnarray}
 \left.\sigma^{(S)}v\right|_{\gamma\gamma}
 &=&
 2\left[{\bf \Gamma}_{\gamma\gamma}\right]_{11}~|d_{21}(mv^2/4)|^2~,
\end{eqnarray}
where the partial absorptive part to two photons ${\bf
\Gamma}_{\gamma\gamma}$ is given in Eq.~(\ref{WG}) for the triplet
EWIMP case and in Eq.~(\ref{HG}) for the doublet EWIMP case,
respectively.  Here we show only the result in the triplet EWIMP case
for simplicity. The coefficient $d_{21}$ is obtained by solving the
Schroedinger equation (\ref{Schrodinger}). When we expand the solution
by the potential term ${\bf V}(r)$, $\left[{\bf g}_>(r)\right]_{21}$
is obtained at leading order as
\begin{eqnarray}
 \left[{\bf g}_>(r)\right]_{21}
 =
 d_{21}(E)~e^{i\sqrt{mE}r}
 \simeq
 \frac{-m\sqrt{2}\alpha_2}{m_W + \sqrt{2m\delta m}}~
 e^{i\sqrt{mE}r}~.
 \label{purturbative d}
\end{eqnarray}
Here, we take $E<2\delta m$ and $m
\gg m_W$. Thus, the annihilation cross section to two photons is given as
\begin{eqnarray}
 \left.\sigma^{(S)}v\right|_{\gamma\gamma}
 \simeq
 \frac{4\pi\alpha^2\alpha_2^2}{m_W^2}
 \left(
  1 + \sqrt{\frac{2m\delta m}{m_W^2}}
 \right)^{-2}~.
 \label{perturbative CS}
\end{eqnarray}
This agrees with the result obtained in the full one-loop calculation in
a heavy EWIMP mass limit \cite{Bergstrom:1997fh}, including the
correction due to the non-vanishing $\delta m$. 

\vspace{0.5cm}
\underline{4. 
Cross section for $\tilde{\chi}^0\tilde{\chi}^0\rightarrow \gamma\gamma$
in a limit of $m\rightarrow\infty$}
\vspace{0.5cm}

When the EWIMP mass $m$ is much heavier than the weak gauge boson masses,
we can not deal with the potential term perturbatively. The equation
(\ref{Schrodinger}) can be solved analytically in the limit of
$m\rightarrow\infty$, and the qualitative behavior of the cross
sections can be discussed. This is because the weak gauge boson masses
$m_W$ and $m_Z$ and the mass difference $\delta m$ can be neglected in
this limit. The Schroedinger equation for ${\bf g}_>(r)$ becomes
\begin{eqnarray}
 -\frac{1}{m}\frac{d^2}{dr^2}{\bf g}_>(r)
 +\frac{1}{r}{\bf U}{\bf g}_>(r)
 =
 E{\bf g}_>(r)~,
\end{eqnarray}
where ${\bf U}$ is the coefficient matrix for the electroweak
potential, and defined by ${\bf U} = [r{\bf V}(r)]_{r \rightarrow
0}$. Because all forces in the potential become Coulomb-type, the
solution of above equation is determined by using the confluent
geometric function as
\begin{eqnarray}
 &~&\qquad\qquad\qquad
 \left[{\bf g}_>(r)\right]_{ij}
 =
 \sum_{i'}
 {\cal O}_{ii'}
 \Lambda_{i'}(r)
 {\cal O}^T_{i'j}~,
 \nonumber \\
 \nonumber \\
 &~&
 \Lambda_i(r)
 =
 \Gamma\left(1 + \frac{i\lambda_i}{2}\sqrt{\frac{m}{E}}\right)
 W_{-\frac{i\lambda_i}{2}\sqrt{\frac{m}{E}},\frac{1}{2}}
 \left(-2i\sqrt{mE}r\right)~.
\end{eqnarray}
The function $\Gamma(z)$ is the Euler's Gamma function and
$W_{\kappa,\mu}(z)$ is the Whittaker function. The matrix ${\cal O}$ is
the diagonalization matrix, ${\cal O}^T{\bf U}{\cal O} = {\rm
diag.}(\lambda_1,\cdots)$, and $\lambda_i$ is the eigenvalue of the
matrix ${\bf U}$. 

The EWIMP annihilation cross section to two photons is then derived as
\begin{eqnarray}
 \sigma^{(S)}v
 &=&
 \frac{\pi\alpha^2}{m^2}
 \left|
  \sum_i
  {\cal O}_{i2}
  \Gamma\left(1 + \frac{i\lambda_i}{2}\sqrt{\frac{m}{E}}\right)
  {\cal O}_{i1}
 \right|^2
 \nonumber \\
 \nonumber \\
 &\simeq&
 \left\{
  \begin{array}{l}
   2.8\times 10^{-5}/vm^2
   \qquad
   ({\rm Triplet~EWIMP})
   \\
   3.2\times 10^{-6}/vm^2
   \qquad
   ({\rm Doublet~EWIMP})
  \end{array}
 \right.~,
\end{eqnarray} 
in the limit of $m\rightarrow\infty$. Here, we use the approximation
$v\ll 1$ to derive the last equation. The cross section behaves as
$\sigma v \sim 1/(vm^2)$, and satisfies the unitarity condition in
Eq.~(\ref{unitarity bound}) as expected.


\vspace{1.0cm}
\lromn 5 \hspace{0.2cm}
{\bf Annihilation Cross Sections and Zero-Energy Resonances}
\vspace{0.5cm}

When the EWIMP mass is heavy enough so that the the effects of the
long-distance force by the electroweak potential cannot be ignored but
not heavy enough to take the limit of $m\rightarrow\infty$, we have
to solve the Schroedinger equation (\ref{Schrodinger}) numerically for
the precise annihilation cross sections. In this section we show some
numerical results, and discuss the behaviors using a toy model.

\vspace{0.5cm}
\underline{1. Numerical result and zero-energy resonances}
\vspace{0.5cm}

\begin{figure}[t]
 \begin{center}
  \includegraphics[height = 4.8cm,clip]{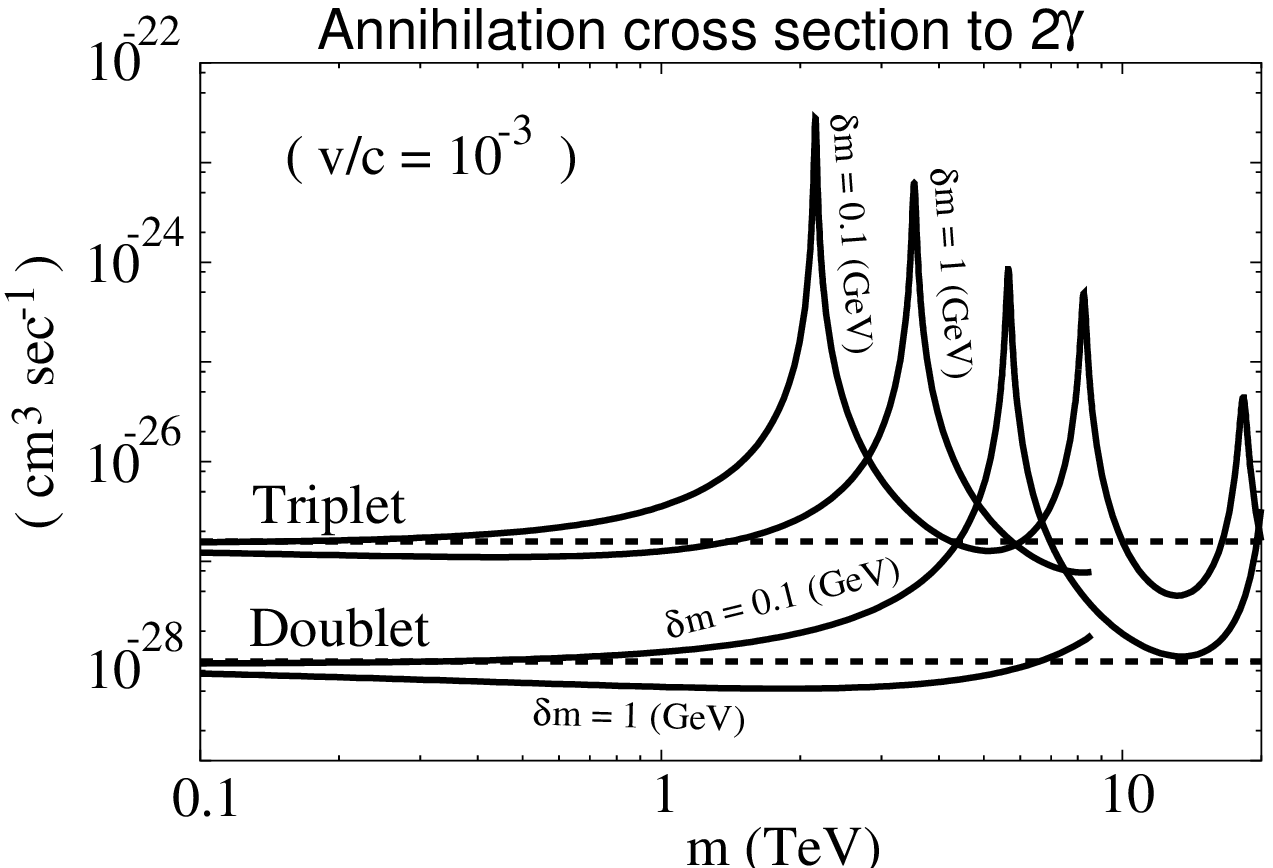}
  ~~
  \includegraphics[height = 4.8cm,clip]{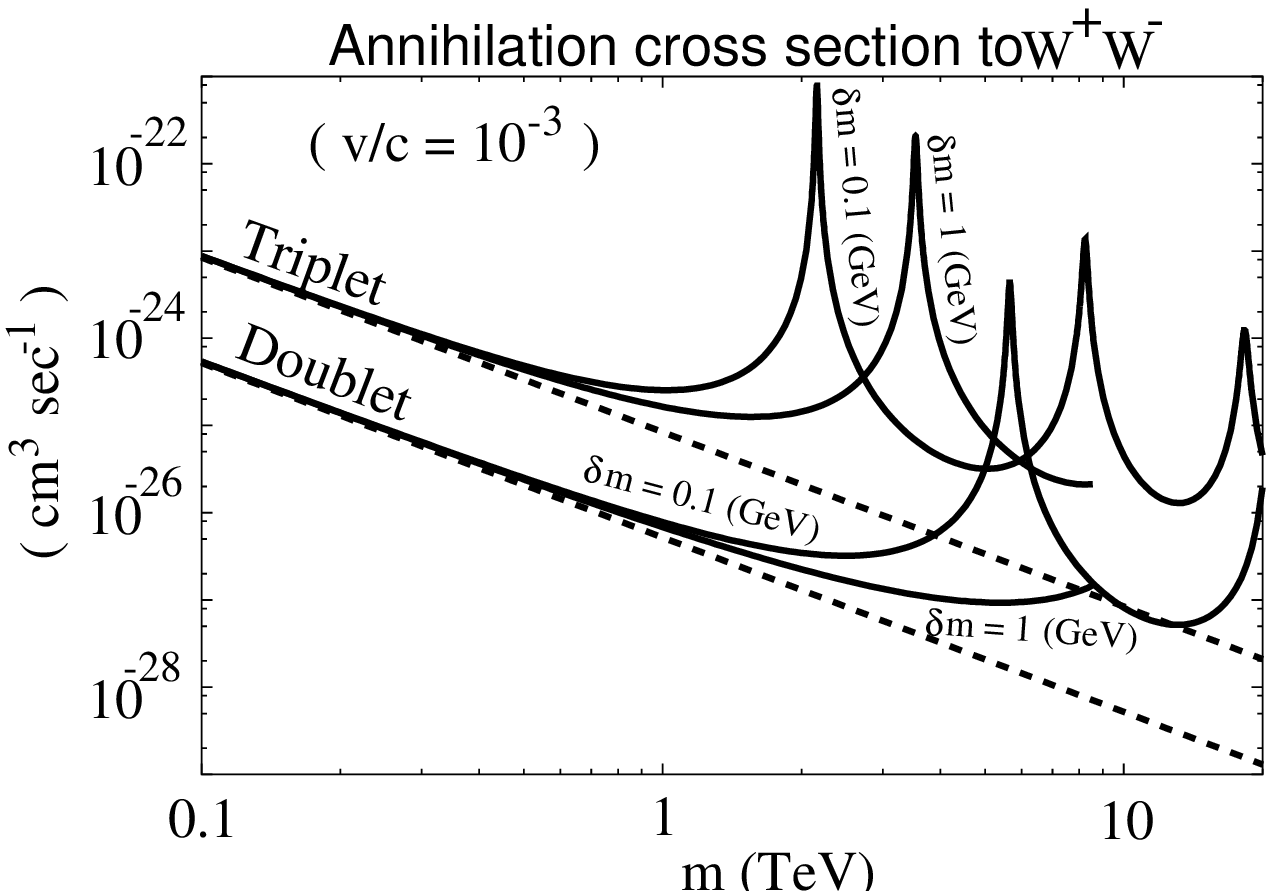}
 \end{center}
 \caption{\small
 Annihilation cross sections ($\sigma v$) to $\gamma\gamma$ and $W^+W^-$ when 
 $\delta m = 0.1, 1$ GeV for both the triplet and the doublet
 EWIMPs. Here, $v/c = 10^{-3}$. The leading-order cross sections in 
 perturbation are also shown for $\delta m = 0$ (broken lines).
 \label{cross section I}
 }
\end{figure}

First, we show numerical results for the triplet and doublet EWIMP
annihilation cross sections to $\gamma\gamma$ and $W^+W^-$. In
Fig.~\ref{cross section I}, the cross sections with some fixed mass
differences ($\delta m = 0.1, 1$ GeV) are shown as functions of $m$. 
We set to the mass difference $\delta m_N$ to be $\delta m_N = 2\delta
m$ for the doublet EWIMP. In
this calculation, the relative velocity of the incident EWIMPs
$v/c$ is taken to be $10^{-3}$, which is the typical velocity of 
dark matter in the galactic halo. We also show the leading-order cross
sections in perturbation as dotted lines. When the EWIMP mass $m$ is
around 100 GeV, the cross sections to $\gamma\gamma$ and $W^+W^-$ are almost
the same as the perturbative ones. However, when $m$ is large enough ($m
\gsim 0.5$ TeV for the triplet EWIMP and $m \gsim 1.5$ TeV for the
doublet EWIMP), the cross sections are significantly enhanced and have
resonance behaviors.

The qualitative behavior of the cross sections, especially around the
resonances, may be understood by using a toy model, in which the
electroweak potential is approximated by a well potential.  Here we
discuss the triplet EWIMP case, and the extension to the doublet case is
straightforward. Taking $c_W = 1$ and $\delta m = 0$ for simplicity, the 
electroweak potential is approximated as  
\begin{eqnarray}
 {\bf V}(r)
 =
 \begin{pmatrix}
  -b_1\alpha_2m_W & -b_1\sqrt{2}\alpha_2m_W
  \\
  -b_1\sqrt{2}\alpha_2m_W & 0
 \end{pmatrix}
 \theta((b_2m_W)^{-1} - r)~,
 \label{toy V}
\end{eqnarray}
where $b_1$ and $b_2$ are numerical constants. By comparing the
annihilation cross sections to $\gamma\gamma$ in this potential and in the
perturbative calculation Eq.~(\ref{perturbative CS}) for small $m$, we
find $b_1 = 8/9$ and $b_2 = 2/3$. Under this potential, the two-body
states of $2\tilde{\chi}^0$ and $\tilde{\chi}^+\tilde{\chi}^-$ have 
attractive and repulsive states, whose potential energies are
$\lambda_\pm = ({\bf V}_{11} \pm \sqrt{{\bf V}_{11}^2 + 4{\bf
V}_{12}^2})/2$ with ${\bf V}_{ij}~(i,j = 1,2)$ elements in ${\bf
V}$. The attractive state is $\cos\theta~\phi_N - \sin\theta~\phi_C$
with $\tan^2\theta = -\lambda_-/\lambda_+$.

By virtue of the approximation, the pair annihilation cross sections
for the triplet EWIMP are obtained analytically,
\begin{eqnarray}
 &&
 \left(\sigma v\right)_{W^+W^-}
 =
 \frac{\pi\alpha_2^2}{9m^2}
 \left(
  |d_{21}|^2 + \sqrt{2}\Re(d_{21}d_{22}^*) + 2|d_{22}|^2
 \right)~,
 \qquad
 \left(\sigma v\right)_{\gamma\gamma}
 =
 \frac{2\pi\alpha^2}{9m^2}|d_{21}|^2~,
 \nonumber \\
 \nonumber \\
 &&
 d_{21}
 =
 \left\{
  ~~
  \sqrt{2}
  \left[
   \cos\left(p_c \sqrt{v^2/v_c^2 + 2}\right)
   -
   i\sqrt{\frac{v^2}{v^2 + 2v_c^2}}
   \sin\left(p_c \sqrt{v^2/v_c^2 + 2}\right)
  \right]^{-1}
 \right.
 \nonumber \\
 &&
 \left.
  \qquad~~~
  -
  \sqrt{2}
  \left[
   \cos\left(p_c \sqrt{v^2/v_c^2 - 1}\right)
   -
   i\sqrt{\frac{v^2}{v^2 - v_c^2}}
   \sin\left(p_c \sqrt{v^2/v_c^2 - 1}\right)
  \right]^{-1}
  ~~
 \right\}~,
 \nonumber \\
 &&
 d_{22}
 =
 \left\{
  \qquad
  \left[
   \cos\left(p_c \sqrt{v^2/v_c^2 + 2}\right)
   -
   i\sqrt{\frac{v^2}{v^2 + 2v_c^2}}
   \sin\left(p_c \sqrt{v^2/v_c^2 + 2}\right)
  \right]^{-1}
 \right.
 \nonumber \\
 &&
 \left.
  \qquad\qquad
  +
  2
  \left[
   \cos\left(p_c \sqrt{v^2/v_c^2 - 1}\right)
   -
   i\sqrt{\frac{v^2}{v^2 - v_c^2}}
   \sin\left(p_c \sqrt{v^2/v_c^2 - 1}\right)
  \right]^{-1}
  ~~
 \right\}~,
\nonumber\\
 \label{well potential model}
\end{eqnarray}
where $p_c$ and $v_c$ are defined by $p_c =
\sqrt{2\alpha_2m/m_W}$ and $v_c = \sqrt{32\alpha_2m_W/9m}$,
respectively.

If the kinetic energy of the EWIMP pair is much larger than the
potential energy ($v\gg v_c$) or the electroweak potential is
point-like ($p_c\ll 1$), the cross sections coincide with the results
in leading-order calculation in perturbation as expected. However,
when $v$ is much smaller than $v_c$, such as in the case of dark
matter in the current universe, the $d_{21}$ and $d_{22}$ become
\begin{eqnarray}
 d_{21}
 &\simeq&
 \sqrt{2}
 \left[
  \cos\sqrt{2}p_c
 \right]^{-1}
 -
 \sqrt{2}
 \left[
  \cosh p_c
 \right]^{-1}~,
 \qquad
\nonumber\\
 d_{22}
 &\simeq&
 \left[
  \cos\sqrt{2}p_c
 \right]^{-1}
 +
 2
 \left[
  \cosh p_c
 \right]^{-1}~.
 \label{coefficients}
\end{eqnarray}
Therefore, when $\sqrt{2}p_c \simeq (2n - 1)\pi/2~(n = 1, 2, \cdots)$,
the coefficients $d_{21}$ and $d_{22}$ are enhanced by several orders
of magnitude. As the result, the cross sections show the resonance
features as in Fig.~\ref{cross section I}. These resonances are called
zero-energy resonances \cite{landau}, because the condition
$\sqrt{2}p_c \simeq (2n - 1)\pi/2$ is nothing but existence of a
bound state with zero binding energy. The bound states consist of
mixtures of pairs of the EWIMPs and the charged partners.

In Fig.~\ref{cross section I}, the first resonance in small $\delta m$
(that is $\delta m \sim 0.1$ GeV) appears at $m\sim 2$ TeV. On the other
hand, the well potential model predicts the first resonance at $m\sim
1.8$ TeV. Thus, the model describes the behavior around the first
resonance well. Also, notice that the cross section to two photons, 
which is induced by the one-loop diagrams in perturbation, is
suppressed only by $\alpha^2/\alpha_2^2$ compared with that to $W^+W^-$
for $p_c\gsim 1$. This behavior is also seen in Fig.~\ref{cross section I}.

On the resonance, the coefficients $d_{21}$ and $d_{22}$ in
Eq.~(\ref{coefficients}) behave as 
\begin{eqnarray}
 d_{21} \simeq i\frac{v_c}{v}~,
 \qquad
 d_{22} \simeq i\frac{\sqrt{2}v_c}{v}~,
\end{eqnarray}
for the small relative velocity $v \ll v_c$. Thus the cross sections
$\sigma v$ are proportional to $v^{-2}$. However, this is not a
signature for breakdown of the unitarity condition in
Eq.~(\ref{unitarity bound}). We find from study in the one-flavor
system under the well potential $V$ that $\sigma v$ would be saturated
by the finite width for the two-body system (that is, the absorptive part)
$\Gamma$ when $v \ll mV\Gamma$, and the unitarity is not broken.

So far we have ignored the mass difference $\delta m$ in Eq.~(\ref{toy
V}). When $\delta m$ is not negligible compared with $\sim \alpha_2
m_W$, the potential energy for the attractive state is reduced as
$|\lambda_-| \simeq 16 \alpha_2 m_W/9- 4\delta m /3 $ for $\delta m
\lsim \alpha_2 m_W$ and $|\lambda_-| \simeq 128 (\alpha_2
m_W)^2/(81\delta m)$ for $\delta m \gsim \alpha_2 m_W$, and the
attractive force becomes weaker. Also, the component of the EWIMP pair
in the attractive state becomes smaller. Then, the zero-energy
resonances move to heavier $m$ for larger $\delta m$, and the EWIMP
annihilation cross sections around the resonances become smaller, as
in Fig.~\ref{cross section I}. This effect is more significant in the
doublet EWIMP case than in the triplet one, since the potential energy
by the weak gauge boson exchanges is smaller.

Now we have discussed importance of the zero-energy resonances for the
EWIMP annihilation cross sections. One might consider that the thermal
relic EWIMP abundance is also modified due to the zero-energy
resonances. The answer to the question is ``No''. The typical relative
velocity is given by $v\sim1/3$ at the freeze out temperature $T\sim
m/20$.  Note that the critical velocity $v_c$ for the triplet EWIMP is
given by $v_c
\sim 2\alpha_2 \sim 0.07$ at the first zero-energy resonance where 
$\sqrt{2}p_c = \pi/2$. The critical velocity for the doublet EWIMP is
smaller than that of the triplet one. Thus the relative velocity at
the freeze-out temperature is not small enough to affect the thermal
relic abundance.

One might have another question relevant to the annihilation in the
universe at very low temperature. Since the cross sections are
proportional to $v^{-2}$ on the zero-energy resonance, the EWIMP
annihilation rate (the annihilation cross section multiplied by the
EWIMP number density $n_{\rm EW}$) becomes much larger
after the freezing-out phenomenon stated above. If the annihilation
rate becomes larger than the Hubble constant $H$ at low temperature,
the EWIMP begins to annihilate again (re-coupling). The
usual calculation of dark matter abundance might be changed.

For studying the possibility, we use the previous toy model for the
triplet EWIMP. The thermal-averaged annihilation cross section
$\langle\sigma v\rangle$ on the first zero-energy resonance at
temperature $T$ is derived as
\begin{eqnarray}
 \langle\sigma v\rangle
 =
 q\left(\frac{\rm 1GeV}{T}\right)~,
 \qquad
 ({\rm in~the~unit~of~GeV}^{-2})~,
 \label{CSlowT}
\end{eqnarray}
where $q \sim 7\times 10^{-9}$. It is
found that the condition of the 're-coupling', $\langle\sigma v\rangle
n_{\rm EW} > H$, attributes to the inequality, 
\begin{eqnarray}
 \frac{\rho_{\rm EW}}{s}
 \geq
 7\times 10^{-8} {\rm [GeV]}~,
\end{eqnarray}
where $\rho_{\rm EW} (\equiv m n_{\rm EW})$ is the EWIMP mass
density and $s$ is the entropy of the universe. The lower bound is
larger than the cosmological observation for dark matter ($\rho/s
\simeq 7\times 10^{-9}$ GeV). Thus the re-coupling does not occur in
our universe.

\vspace{0.5cm}
\underline{2. Fitting functions}
\vspace{0.5cm}

The numerical calculation of the EWIMP annihilation cross sections
takes huge CPU power. It is thus convenient to derive the fitting
functions which reproduce the numerical results with enough
precision. We performed two parameter fitting of the annihilation
cross sections for the doublet and triplet EWIMPs, the EWIMP mass $m$
and the mass difference between the EWIMP and its charged partner $\delta
m$. The velocity dependence of the cross sections is weak except for
very narrow regions in the vicinity of the resonances and ignored. For
the doublet EWIMP, the cross sections further depend on the mass
difference between the EWIMP and its neutral partner $\delta m_N$. We
fix $\delta m_N = 2\delta m$ in the derivation of the fitting
functions for simplicity. This relation is valid in the MSSM when
$\tan\beta$ is large.

We use the fitting function,
\begin{eqnarray}
 {\left.\sigma v(m,\delta m)\right|_{\rm fit}} 
 =
 10^{-28}{\rm[ cm^3~s^{-1}]}\times
 \left[
  \sum_{i,j = 0}^6 a_{ij}
  \left(\frac{m}{\rm 1~TeV}\right)^i
  \left(\ds\frac{\delta m}{\rm 1~GeV}\right)^j
 \right]^{-2}~,
 \label{fitting} 
\end{eqnarray}
in the range $0.2~{\rm TeV} \leq m \leq 10~{\rm TeV}$ and $0.1~{\rm GeV}
\leq \delta m \leq 1~{\rm GeV}$. By performing the least-square method
between the fitting function and the numerical integration, 
the coefficients $a_{ij}$ in each process are obtained.
The numerical values of the coefficients $a_{ij}$ for the
annihilation cross sections to $\gamma\gamma$ and $W^+W^-$ in the triplet
EWIMP case and $\gamma\gamma$, $W^+W^-$ and $Z^0Z^0$ in the doublet EWIMP
case are given in Table 1. Other annihilation cross sections are
evaluated from the cross sections to two photons as
\begin{eqnarray}
 \sigma v(\tilde{\chi}^0\tilde{\chi}^0\rightarrow\gamma Z^0)
 &=&
 \sigma v(\tilde{\chi}^0\tilde{\chi}^0\rightarrow \gamma\gamma)
 \times
 \frac{2 \alpha_2 c_W^2}{\alpha}~,
 \nonumber\\ 
 \sigma v(\tilde{\chi}^0\tilde{\chi}^0\rightarrow Z^0Z^0)
 &=&
 \sigma v(\tilde{\chi}^0\tilde{\chi}^0\rightarrow \gamma\gamma)
 \times
 \frac{\alpha_2^2 c_W^4}{\alpha^2}~,
\end{eqnarray}
in the triplet EWIMP case, and
\begin{eqnarray}
 \sigma v(\tilde{\chi}^0\tilde{\chi}^0\rightarrow \gamma Z^0)
 &=&
 \sigma v(\tilde{\chi}^0\tilde{\chi}^0\rightarrow \gamma\gamma)
 \times
 \frac{\alpha_2 (1-2s_W^2)^2}{2\alpha c_W^2}~,
\end{eqnarray}
in the doublet EWIMP case. This is because these processes are induced 
via the transition of the EWIMP pair to the charged partner pair.

\vspace{0.5cm}

\begin{table}
\begin{center}
{\scriptsize
(Triplet EWIMP, to $\gamma\gamma$)\\
\begin{tabular}{|c||r|r|r|r|r|r|r|}
 \hline
 $a_{ij}$ & j = 0~~~~~~& j = 1~~~~~~ & j = 2~~~~~~ & j = 3~~~~~~ & j = 4~~~~~~ & j = 5~~~~~~ & j = 6~~~~~~ \\
 \hline
 i = 0 &    2.56521E$-$1 &    2.88649E$-$2 &    1.24874E$-$1 & $-$4.97529E$-$1 &    8.48760E$-$1 & $-$6.97638E$-$1 &    2.23588E$-$1 \\
 \hline
 i = 1 & $-$1.10191E$-$1 &    8.27605E$-$1 & $-$3.13417E$+$0 &    7.52692E$+$0 & $-$1.02429E$+$1 &    7.27770E$+$0 & $-$2.09385E$+$0 \\
 \hline
 i = 2 & $-$1.59340E$-$2 & $-$4.83605E$-$1 &    2.29288E$+$0 & $-$5.90629E$+$0 &    8.31712E$+$0 & $-$6.03157E$+$0 &    1.75985E$+$0 \\
 \hline
 i = 3 & $-$8.78312E$-$3 &    2.29927E$-$1 & $-$1.11637E$+$0 &    2.82055E$+$0 & $-$3.86927E$+$0 &    2.73075E$+$0 & $-$7.76046E$-$1 \\
 \hline
 i = 4 &    4.66594E$-$3 & $-$5.84155E$-$2 &    2.72064E$-$1 & $-$6.65170E$-$1 &    8.86046E$-$1 & $-$6.08312E$-$1 &    1.68406E$-$1 \\
 \hline
 i = 5 & $-$5.44363E$-$4 &    6.00290E$-$3 & $-$2.73828E$-$2 &    6.56610E$-$2 & $-$8.58639E$-$2 &    5.79081E$-$2 & $-$1.57571E$-$2 \\
 \hline
 i = 6 &    1.95407E$-$5 & $-$2.10046E$-$4 &    9.48764E$-$4 & $-$2.24620E$-$3 &    2.89895E$-$3 & $-$1.92989E$-$3 &    5.18540E$-$4 \\
 \hline
\end{tabular}
}

{\scriptsize
(Triplet EWIMP, to $W^+W^-$)\\
\begin{tabular}{|c||r|r|r|r|r|r|r|}
 \hline
 $a_{ij}$ & j = 0~~~~~~& j = 1~~~~~~ & j = 2~~~~~~ & j = 3~~~~~~ & j = 4~~~~~~ & j = 5~~~~~~ & j = 6~~~~~~ \\
 \hline
 i = 0 & $-$3.36985E$-$3 &    1.23215E$-$2 & $-$4.08164E$-$2 &    5.96407E$-$2 & $-$2.63576E$-$2 & $-$1.42661E$-$2 &    1.13721E$-$2 \\
 \hline
 i = 1 &    5.61768E$-$2 & $-$6.02547E$-$2 &    1.31888E$-$1 & $-$3.54816E$-$2 & $-$2.97169E$-$1 &    4.06116E$-$1 & $-$1.57962E$-$1 \\
 \hline
 i = 2 & $-$4.18499E$-$2 &    8.09353E$-$2 & $-$1.05570E$-$1 & $-$1.60918E$-$1 &    6.66742E$-$1 & $-$7.10708E$-$1 &    2.52266E$-$1 \\
 \hline
 i = 3 &    6.54865E$-$3 & $-$4.19556E$-$4 & $-$8.61235E$-$2 &    3.59611E$-$1 & $-$6.42137E$-$1 &    5.31727E$-$1 & $-$1.66897E$-$1 \\
 \hline
 i = 4 &    1.43330E$-$4 & $-$6.78893E$-$3 &    4.15698E$-$2 & $-$1.20926E$-$1 &    1.83077E$-$1 & $-$1.37981E$-$1 &    4.08085E$-$2 \\
 \hline
 i = 5 & $-$8.33119E$-$5 &    1.03757E$-$3 & $-$5.22051E$-$3 &    1.36654E$-$2 & $-$1.93028E$-$2 &    1.38678E$-$2 & $-$3.96380E$-$3 \\
 \hline
 i = 6 &    4.08625E$-$6 & $-$4.38210E$-$5 &    2.04733E$-$4 & $-$5.07069E$-$4 &    6.86353E$-$4 & $-$4.77300E$-$4 &    1.33068E$-$4 \\
 \hline
\end{tabular}
}

{\scriptsize
(Doublet EWIMP, to $\gamma\gamma$)\\
\begin{tabular}{|c||r|r|r|r|r|r|r|}
 \hline
 $a_{ij}$ & j = 0~~~~~~& j = 1~~~~~~ & j = 2~~~~~~ & j = 3~~~~~~ & j = 4~~~~~~ & j = 5~~~~~~ & j = 6~~~~~~ \\
 \hline
 i = 0 &    1.04998E$+$0 & $-$1.92220E$-$1 &    2.60536E$+$0 & $-$8.25766E$+$0 &    1.24256E$+$1 & $-$9.11921E$+$0 &    2.61707E$+$0 \\
 \hline
 i = 1 & $-$4.17041E$-$1 &    4.24814E$+$0 & $-$1.96145E$+$1 &    5.23145E$+$1 & $-$7.39832E$+$1 &    5.28533E$+$1 & $-$1.49773E$+$1 \\
 \hline
 i = 2 &    1.03008E$-$1 & $-$1.93097E$+$0 &    1.05626E$+$1 & $-$2.97310E$+$1 &    4.29137E$+$1 & $-$3.10093E$+$1 &    8.82103E$+$0 \\
 \hline
 i = 3 & $-$2.42293E$-$2 &    4.15616E$-$1 & $-$2.28085E$+$0 &    6.20020E$+$0 & $-$8.49779E$+$0 &    5.80186E$+$0 & $-$1.52947E$+$0 \\
 \hline
 i = 4 &    2.24803E$-$3 & $-$3.43865E$-$2 &    1.66100E$-$1 & $-$3.65744E$-$1 &    3.37953E$-$1 & $-$9.30267E$-$2 & $-$2.82121E$-$2 \\
 \hline
 i = 5 & $-$6.51687E$-$5 &    6.85671E$-$4 & $-$2.51709E$-$4 & $-$1.22816E$-$2 &    4.22829E$-$2 & $-$5.14471E$-$2 &    2.24206E$-$2 \\
 \hline
 i = 6 & $-$1.47604E$-$7 &    1.74390E$-$5 & $-$2.40349E$-$4 &    1.19010E$-$3 & $-$2.71981E$-$3 &    2.87046E$-$3 & $-$1.16460E$-$3 \\
 \hline
\end{tabular}
}

{\scriptsize
(Doublet EWIMP, to $W^+W^-$)\\
\begin{tabular}{|c||r|r|r|r|r|r|r|}
 \hline
 $a_{ij}$ & j = 0~~~~~~& j = 1~~~~~~ & j = 2~~~~~~ & j = 3~~~~~~ & j = 4~~~~~~ & j = 5~~~~~~ & j = 6~~~~~~ \\
 \hline
 i = 0 & $-$6.51180E$-$3 &    2.79481E$-$2 & $-$5.26301E$-$2 &    4.49224E$-$2 &    9.13789E$-$3 & $-$4.38570E$-$2 &    2.11067E$-$2 \\
 \hline
 i = 1 &    1.71330E$-$1 & $-$1.78724E$-$1 &    5.13363E$-$1 & $-$8.98539E$-$1 &    8.56009E$-$1 & $-$3.75230E$-$1 &    4.87411E$-$2 \\
 \hline
 i = 2 & $-$5.42160E$-$2 &    1.85541E$-$1 & $-$5.40450E$-$1 &    9.57081E$-$1 & $-$8.89508E$-$1 &    3.57243E$-$1 & $-$3.19227E$-$2 \\
 \hline
 i = 3 &    2.44545E$-$3 & $-$9.05849E$-$4 & $-$3.64007E$-$2 &    1.57625E$-$1 & $-$3.13715E$-$1 &    2.94701E$-$1 & $-$1.03643E$-$1 \\
 \hline
 i = 4 &    2.49469E$-$4 & $-$3.59210E$-$3 &    2.13727E$-$2 & $-$6.52043E$-$2 &    1.08466E$-$1 & $-$9.07532E$-$2 &    2.95480E$-$2 \\
 \hline
 i = 5 & $-$2.49263E$-$5 &    3.33127E$-$4 & $-$2.02985E$-$3 &    6.39148E$-$3 & $-$1.07614E$-$2 &    9.00477E$-$3 & $-$2.92073E$-$3 \\
 \hline
 i = 6 &    6.51335E$-$7 & $-$9.63636E$-$6 &    6.24283E$-$5 & $-$2.04190E$-$4 &    3.50050E$-$4 & $-$2.95151E$-$4 &    9.60757E$-$5 \\
 \hline
\end{tabular}
}

{\scriptsize
(Doublet EWIMP, to $Z^0Z^0$)\\
\begin{tabular}{|c||r|r|r|r|r|r|r|}
 \hline
 $a_{ij}$ & j = 0~~~~~~& j = 1~~~~~~ & j = 2~~~~~~ & j = 3~~~~~~ & j = 4~~~~~~ & j = 5~~~~~~ & j = 6~~~~~~ \\
 \hline
 i = 0 & $-$4.38874E$-$5 & $-$9.72321E$-$3 &    7.53513E$-$2 & $-$2.21045E$-$1 &    3.21256E$-$1 & $-$2.32073E$-$1 &    6.64397E$-$2 \\
 \hline
 i = 1 &    1.44964E$-$1 &    8.69220E$-$2 & $-$5.06389E$-$1 &    1.34705E$+$0 & $-$1.86944E$+$0 &    1.31757E$+$0 & $-$3.72003E$-$1 \\
 \hline
 i = 2 & $-$6.16430E$-$3 & $-$1.64703E$-$1 &    8.62852E$-$1 & $-$2.17291E$+$0 &    2.92730E$+$0 & $-$2.02460E$+$0 &    5.63888E$-$1 \\
 \hline
 i = 3 & $-$1.36278E$-$2 &    1.29107E$-$1 & $-$5.70725E$-$1 &    1.35708E$+$0 & $-$1.77804E$+$0 &    1.20897E$+$0 & $-$3.32609E$-$1 \\
 \hline
 i = 4 &    2.05191E$-$3 & $-$1.94524E$-$2 &    8.76460E$-$2 & $-$2.13319E$-$1 &    2.86459E$-$1 & $-$1.99525E$-$1 &    5.61327E$-$2 \\
 \hline
 i = 5 & $-$1.06900E$-$4 &    1.07990E$-$3 & $-$5.08739E$-$3 &    1.28877E$-$2 & $-$1.79673E$-$2 &    1.29591E$-$2 & $-$3.76293E$-$3 \\
 \hline
 i = 6 &    1.93736E$-$6 & $-$2.09370E$-$5 &    1.03637E$-$4 & $-$2.74183E$-$4 &    3.97811E$-$4 & $-$2.97743E$-$4 &    8.93856E$-$5 \\
 \hline
\end{tabular}
}
\end{center}
\caption{\small Coefficients of the fitting function Eq.~(\ref{fitting})
 for the EWIMP annihilation cross sections to $\gamma\gamma$ and $W^+W^-$ in  
 the triplet case, and to $\gamma\gamma$, $W^+W^-$ and $Z^0Z^0$ in the doublet
 case.} 
\label{table1}
\end{table}

\vspace{0.5cm}
\underline{3. Annihilation cross sections in the MSSM parameters}
\vspace{0.5cm}

By using above fitting functions, we scan the
annihilation cross sections on the MSSM parameters. The input parameters
are the Bino mass $M_1$, the Wino mass $M_2$, the Higgsino mass $\mu$,
and tan$\beta$. We assume that the other supersymmetric scalar particles,
sleptons and squarks, are heavy enough, and neglect the contributions to
the cross sections. Furthermore we consider two situations, in which
relations between $M_1$ and $M_2$ are different. The first one
is the GUT relation $M_1 = 0.5M_2$, which is frequently assumed in the minimal
supergravity scenario. Another one is $M_1 = 3M_2$, which is predicted in the
anomaly mediated supersymmetry breaking scenario. The lightest
neutralino mass $m$ and the mass difference between the neutralino and
the chargino $\delta m$ in these two cases are given in
Fig.~\ref{massdiff1} and Fig.~\ref{massdiff2} in Sec. \lromn{2}.

The non-perturbative effects due to the existence of the resonances
are important for the calculation of the annihilation cross sections
when the neutralino is Wino- or Higgsino-like and heavy enough. On the
other hand, the leading-order calculation is precise enough in other
regions of the parameters. We thus match the fitting functions derived
in the previous section to the leading-order cross sections in
perturbation at $m = 250$ GeV and $\delta m = 1$ GeV with $|Z_{12}|^2
\geq 0.9$ for the Wino-like and $|Z_{13}|^2 + |Z_{14}|^2 \geq 0.9$ for
the Higgsino-like neutralino. Here, $Z_{ij}$ is defined in
Eq.~(\ref{neutralino comp}).

\begin{figure}[p]
 \begin{center}
  \includegraphics[height = 9.5cm,clip]{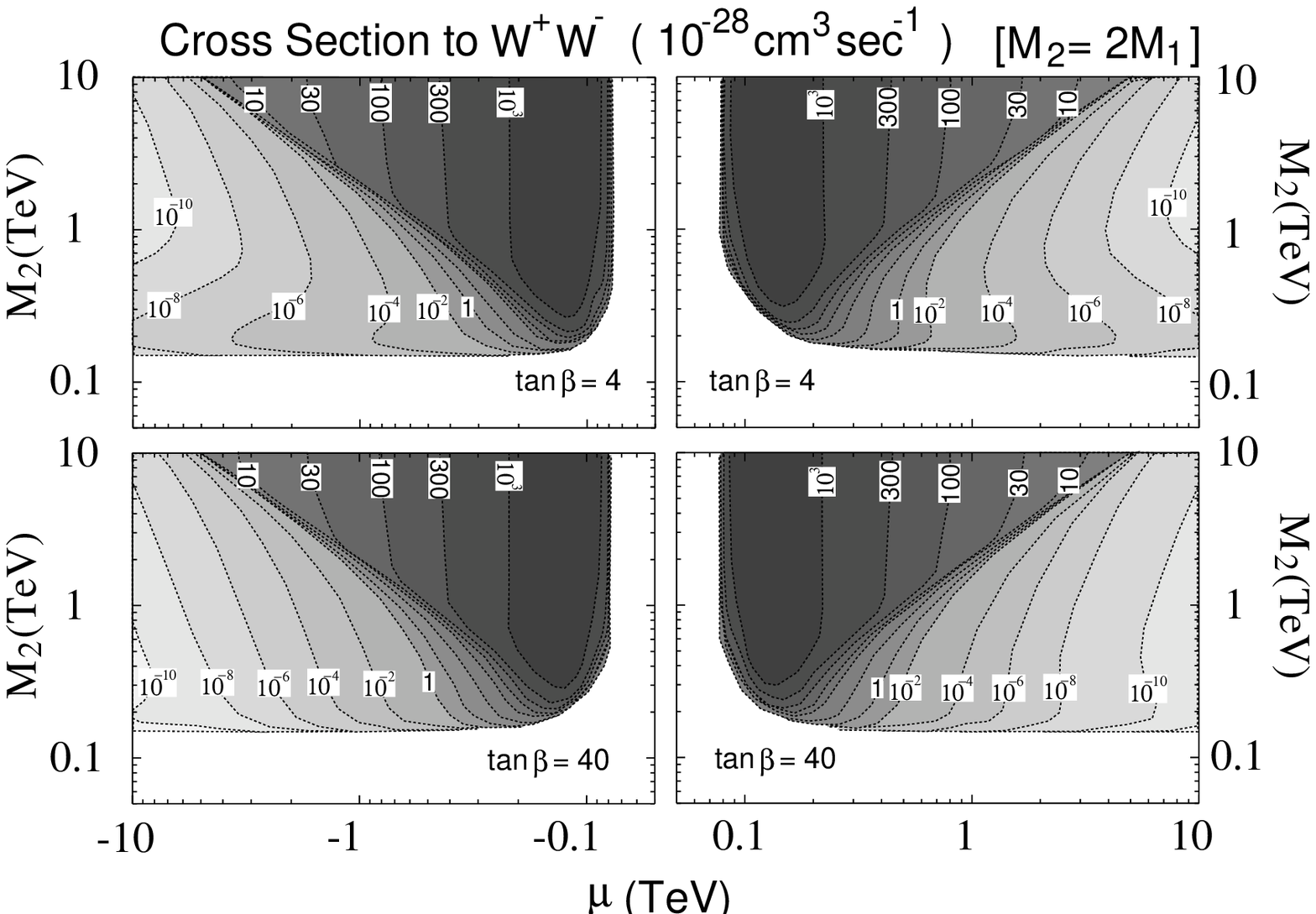}
  \\
  ~
  \\
  \includegraphics[height = 9.5cm,clip]{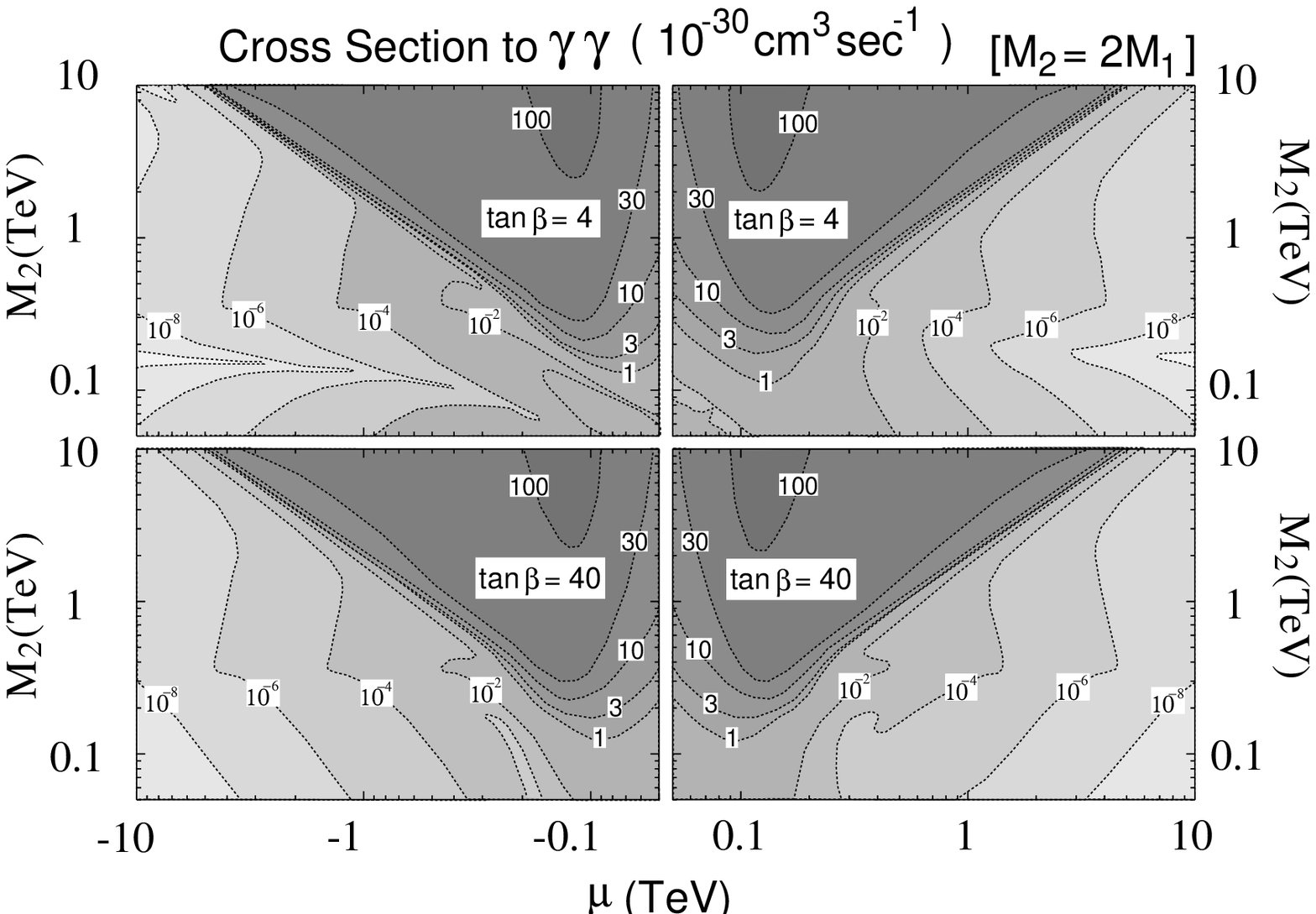}
 \end{center}
 \caption{\small
 Contour maps of the neutralino annihilation cross sections ($\sigma v$)
 to $W^+W^-$
 (four top figures) and $\gamma\gamma$ (four bottom figures) in ($M_2,\mu$)
 planes with $\tan\beta = 4, 40$ in the MSSM. 
 $M_1 = 0.5M_2$ is assumed.
 \label{CSI}
 }
\end{figure}

\begin{figure}[p]
 \begin{center}
  \includegraphics[height = 9.5cm,clip]{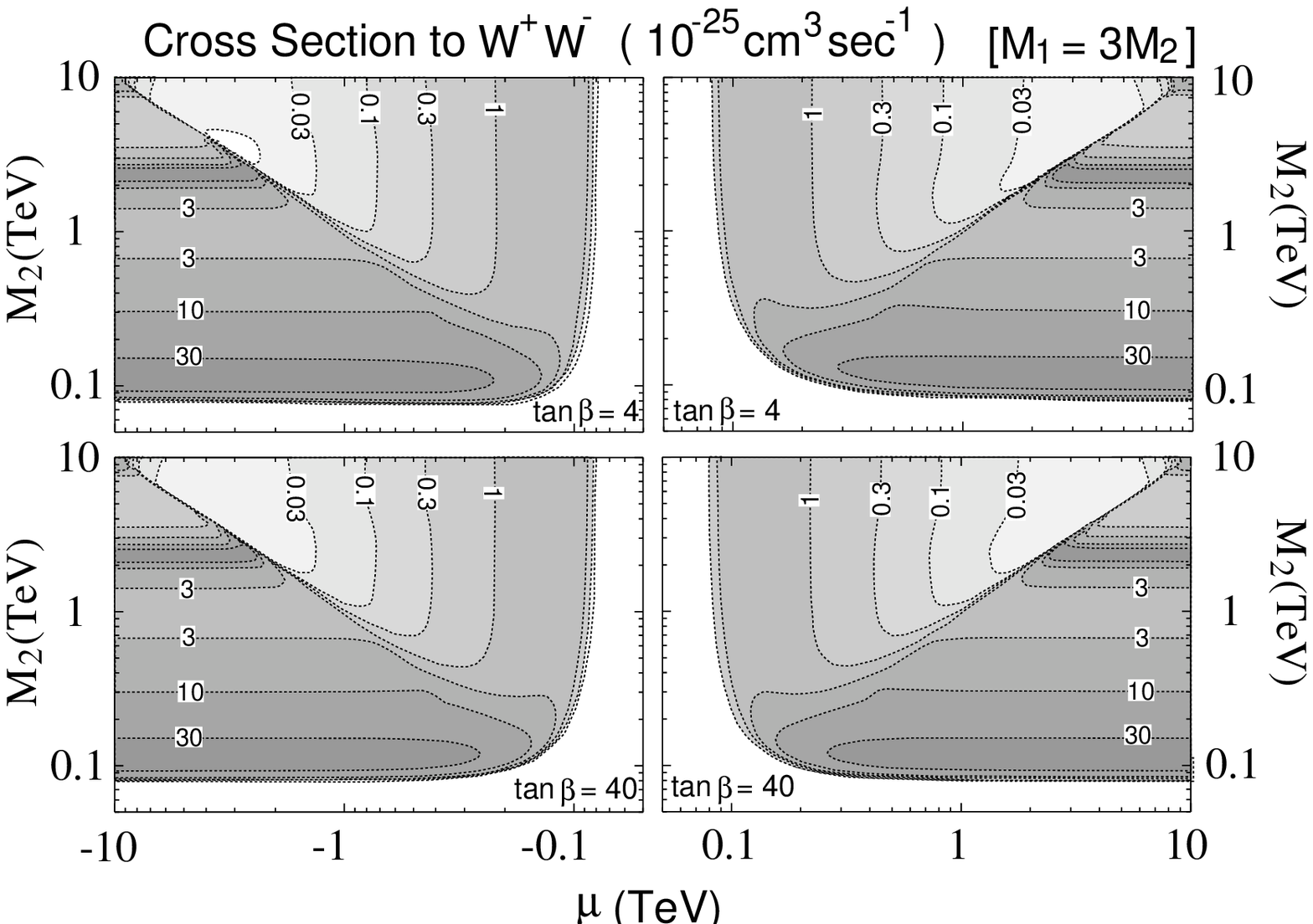}
  \\
  ~
  \\
  \includegraphics[height = 9.5cm,clip]{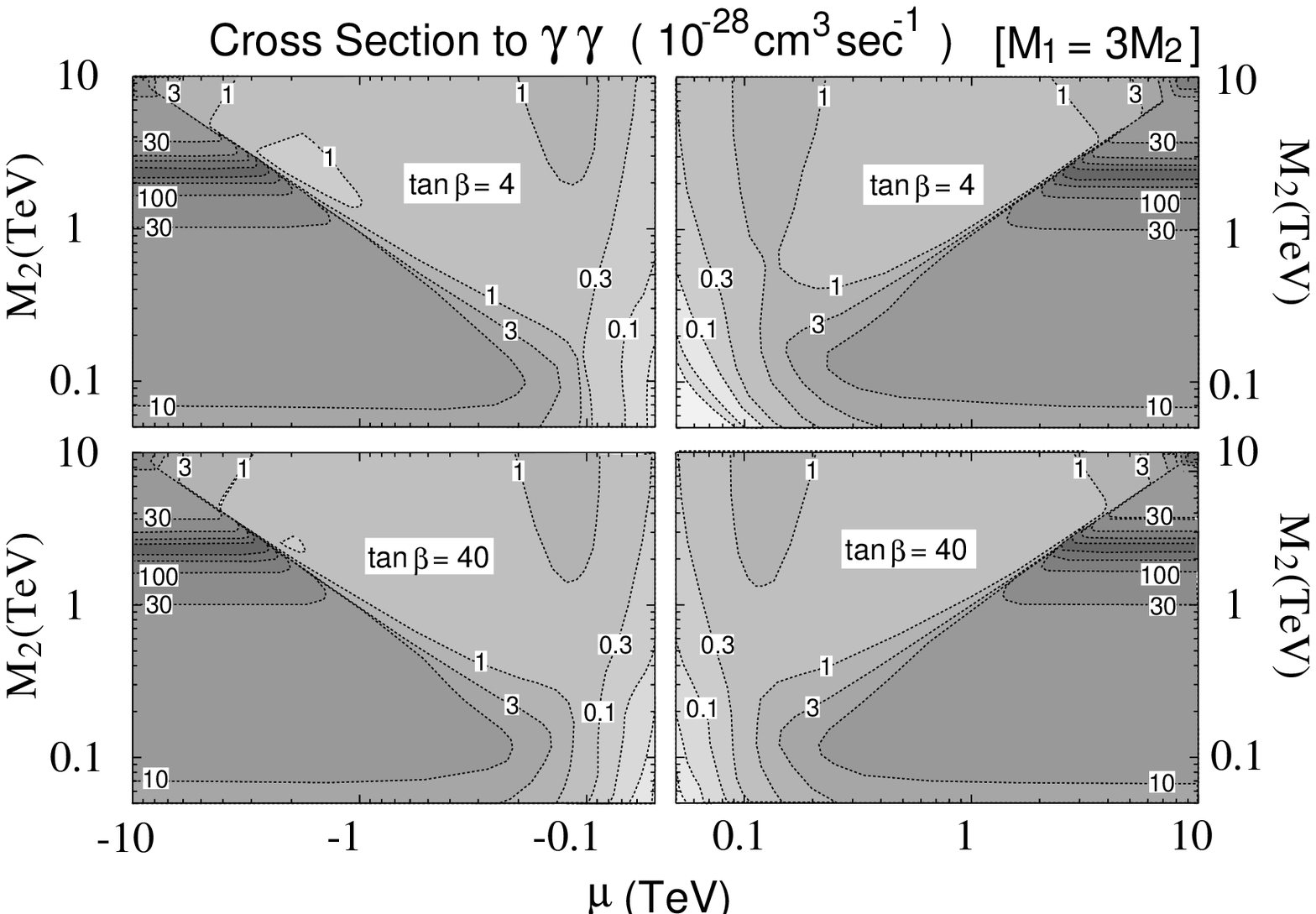}
 \end{center}
 \caption{\small
 Contour maps of the neutralino annihilation cross sections  ($\sigma v$) to $W^+W^-$
 (four top figures) and $\gamma\gamma$ (four bottom figures) in ($M_2,\mu$)
 planes with  $\tan\beta = 4, 40$ in the MSSM. 
 $M_1 = 3M_2$ is assumed.
 \label{CSII}
 }
\end{figure}

The results are shown in Fig.~\ref{CSI} and Fig.~\ref{CSII}, which are
contour plots of the cross sections to $W^+W^-$ and $\gamma\gamma$. In
Fig.~\ref{CSI}, the relation $M_1 = 0.5M_2$ is imposed. The regions
$|\mu|\lsim 0.5M_2$ correspond to the Higgsino-like
regions. As in Fig.~\ref{massdiff1}, the mass difference between the
neutralino and the chargino is larger than 1 GeV in most of the parameter
space. As the result, the first resonance appears at $|\mu|\sim $10
TeV, which is out of the range of these figures. In the  regions with
$0.5M_2\lsim |\mu|$  the LSP is the Bino-like neutralino.

The relation $M_1 = 3M_2$ is imposed in Fig.~\ref{CSII}. In addition
to the Higgsino-like regions, which appears in the same place as ones in
Fig.~\ref{CSI}, the Wino-like regions also appear in the regions $M_2\lsim
|\mu|$.  The first zero-energy resonance of the Wino-like
neutralino is shown up at $M_2\sim 2$ TeV.


\begin{figure}[p]
 \begin{center}
  \includegraphics[height = 9.5cm,clip]{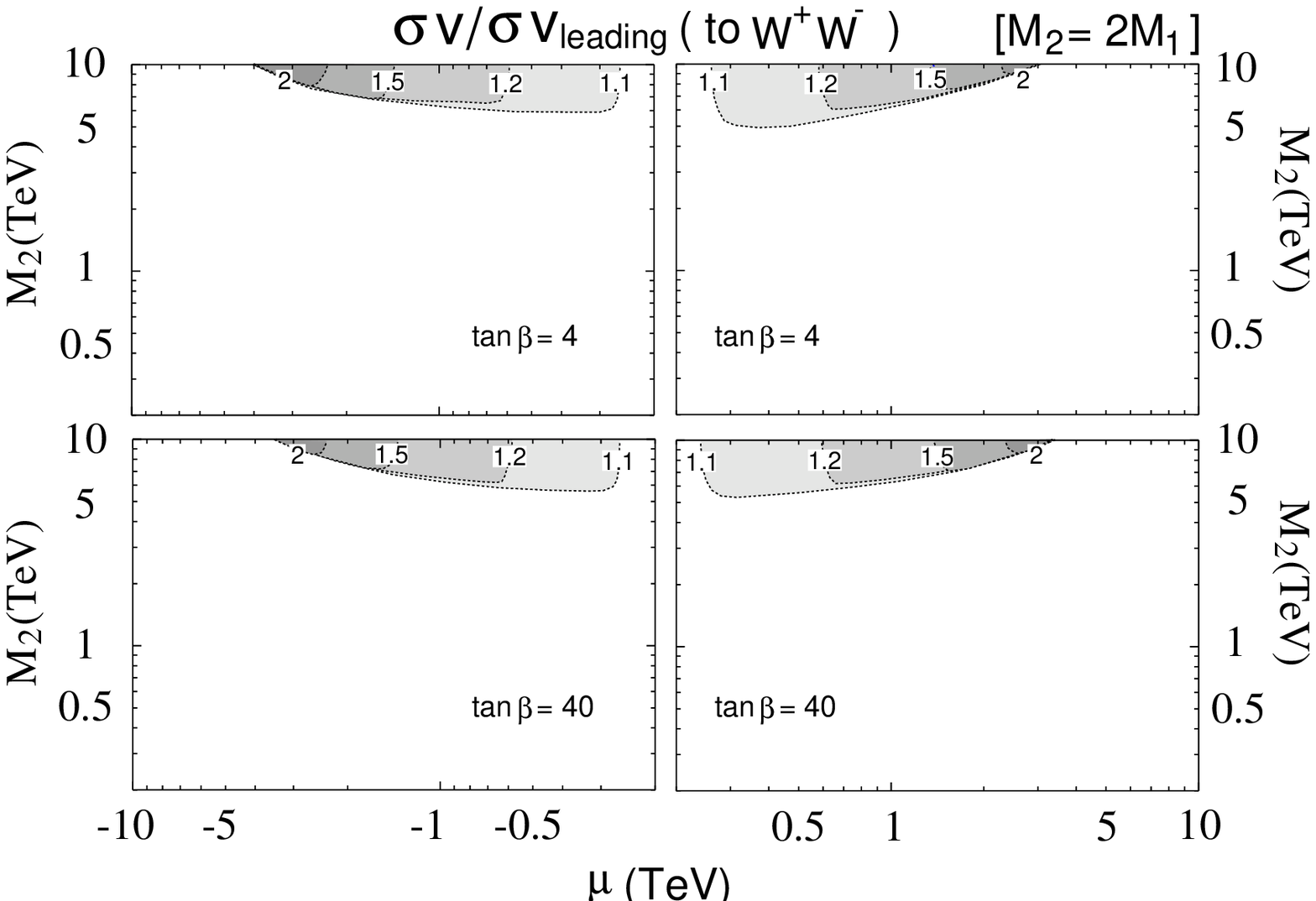}
  \\
  ~
  \\
  \includegraphics[height = 9.5cm,clip]{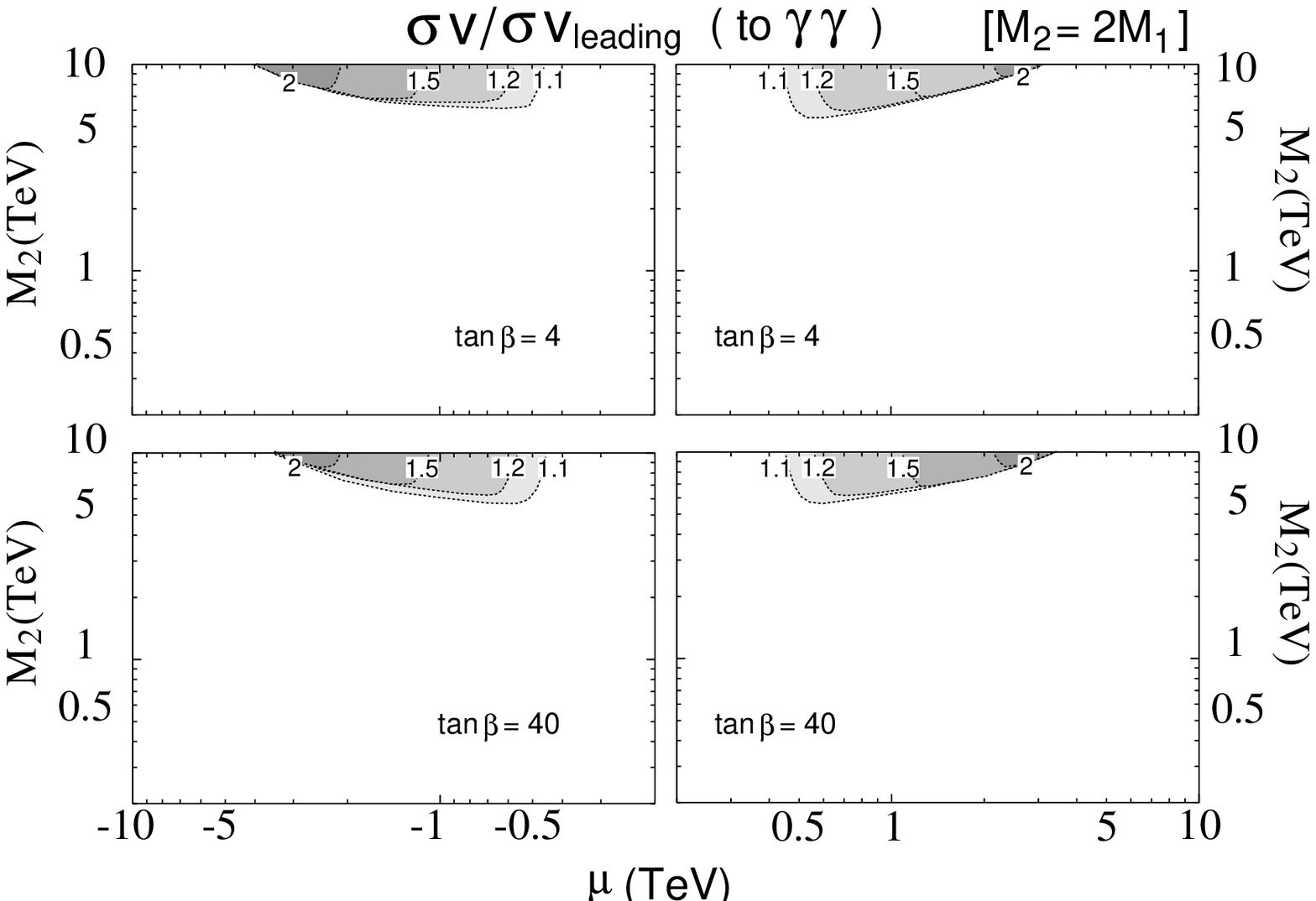}
 \end{center}
 \caption{\small
 Ratios of the cross sections including the non-perturbative effects and
 the leading-order ones in perturbation, $\sigma/\sigma_{\rm leading}$,
 in the MSSM. Figures are shown as contour maps in ($M_2,\mu$)
 planes. All parameters needed for depicting the lines are the same as
 those in  Fig.~\ref{CSI}. 
 \label{CSrI}
 }
\end{figure}

\begin{figure}[p]
 \begin{center}
  \includegraphics[height = 9.5cm,clip]{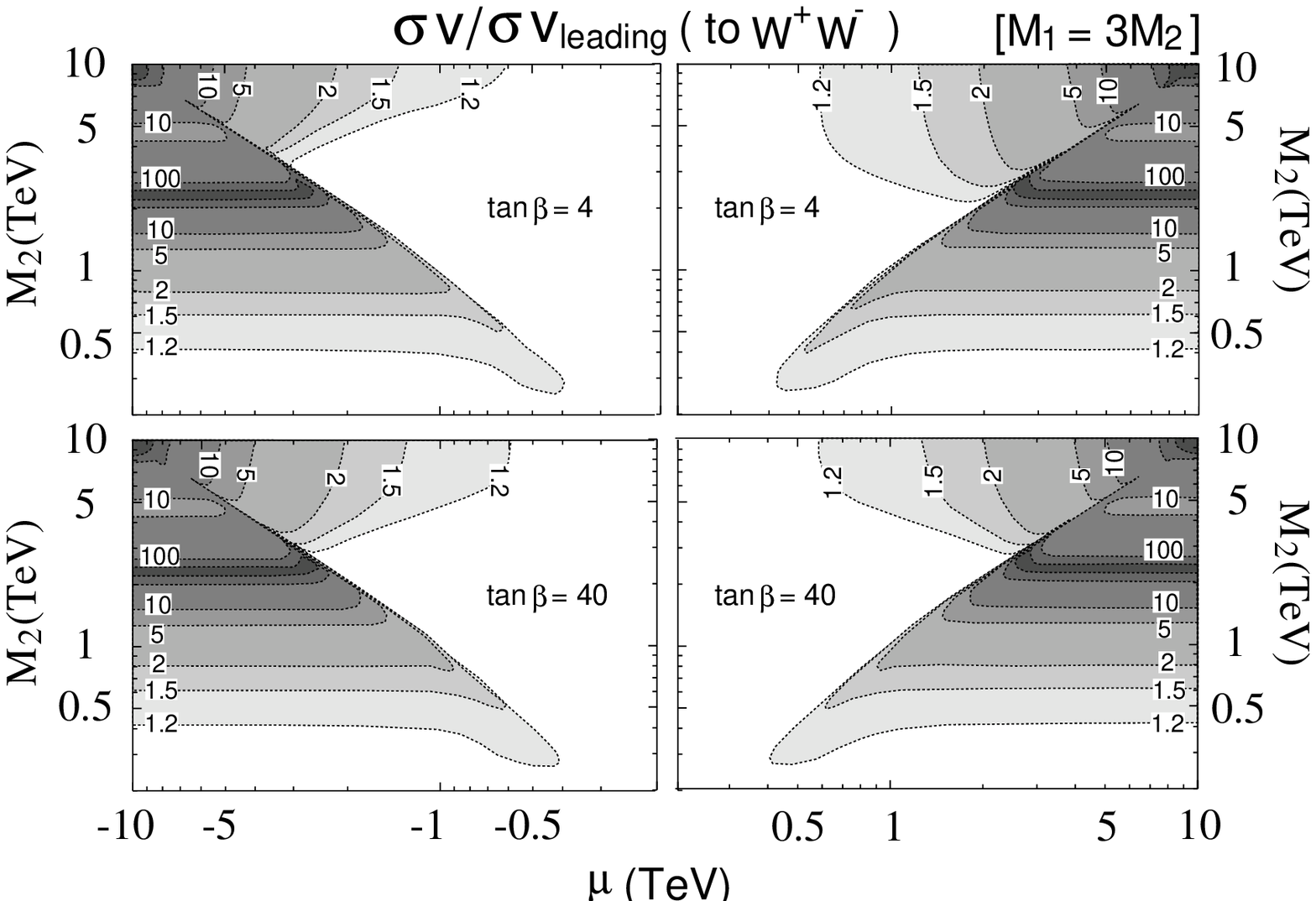}
  \\
  ~
  \\
  \includegraphics[height = 9.5cm,clip]{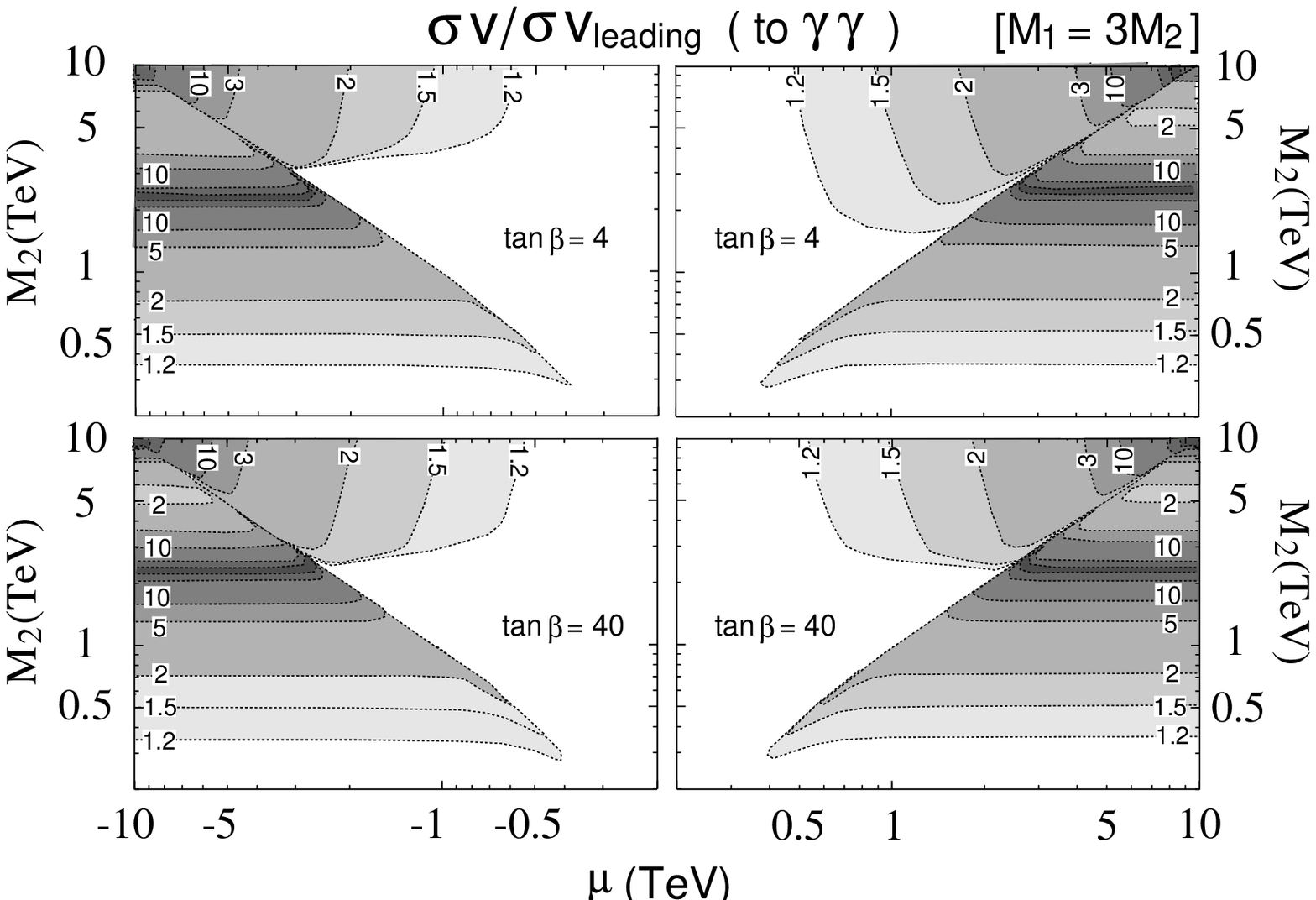}
 \end{center}
 \caption{\small
 Ratios of the cross sections including the non-perturbative effects and
 the leading-order ones in perturbation, $\sigma/\sigma_{\rm leading}$,
 in the MSSM. Figures are shown as contour maps in ($M_2,\mu$)
 planes. All parameters needed for depicting the lines are the same as
 those in Fig.~\ref{CSII}.
 \label{CSrII}
 }
\end{figure}

We also show the ratios of the annihilation cross sections including
the effects of the resonances and the leading-order ones in perturbation
in Fig.~\ref{CSrI} and Fig.~\ref{CSrII}. All parameters for depicting
these figures are the same as the those for Fig.~\ref{CSI} and
Fig.~\ref{CSII}. Huge enhancements are found in the vicinities of the
zero-energy resonances.  Furthermore, the cross sections are enhanced
by factors even for $m\gsim$500GeV when the lightest neutralino is
Wino-like. Thus, the non-perturbative effect is important to obtain
the precise annihilation cross sections.  The ratios in the
Higgsino-like regions in Fig.~\ref{CSrII} are larger than that in
Fig.~\ref{CSrI}. This is because the mass difference $\delta m$ is
smaller in Fig.~\ref{CSrII} than in Fig.~\ref{CSrI}.  The
non-perturbative effects should also be included for the Higgsino-like
neutralino with $m\gsim 500$ GeV when the accurate cross sections are
required.



\vspace{1.0cm}
\lromn 6 \hspace{0.2cm}
{\bf Gamma Ray Flux from the Galactic Center}
\vspace{0.5cm}

The enhancement of the dark matter annihilation cross sections has significant
implication for indirect searches for dark matter using cosmic rays.
In this section, we discuss the searches for gamma rays resulting from
the dark matter annihilation in the galactic center and the future prospects.

\vspace{0.5cm}
\underline{1. Flux formula}
\vspace{0.5cm}

The spectrum of gamma rays from the dark matter annihilation consists of
two components. One is the line gamma rays and the other is the
continuum gamma rays. The line gamma rays are produced by the radiative
processes such as the dark matter annihilations to $\gamma\gamma$ and
$Z\gamma$ \cite{Bergstrom:1997fh}. Since the dark matter is
non-relativistic in the galactic 
halo, the resulting spectrum is monochromatic. The signal is robust
for the dark matter search, because the diffused gamma-ray background
from the astrophysical sources has a continuous energy
spectrum. 

The continuum gamma ray signal come from jets in the dark matter
annihilation. For example, the dark matter annihilates to $W$ bosons,
the $W$ bosons fragment into $\pi$ mesons and finally $\pi^0$ mesons
decay to $\gamma\gamma$. The energy spectrum thus becomes
continuous. The observation of the continuum gamma rays may also
constrain the properties of dark matter if the astrophysical
background is understood well.

The gamma ray flux from the dark matter annihilation, ${\cal
F}_\gamma(E)$, is given by 
\begin{eqnarray}
 \frac{d{\cal F}_\gamma(\theta,E)}{d\Omega dE}
 =
 \frac{1}{4\pi m^2}\sum_f\frac{dN_f^{(\gamma)}}{dE}
 \frac{\langle \sigma v\rangle_f}{2}
 \int_{\rm line~of~sight}dl(\theta)~\rho^2(l)~,
 \label{flux formula 1}
\end{eqnarray}
where $\theta$ is the angle between the direction of the galactic center
and that of observation. The function $N_f^{(\gamma)}(E)$ is the number
of photons with energy $E$ in the fragmentation of the final state $f$,
and $\left<\sigma 
v\right>$ is the dark matter annihilation cross section averaged with
respect to the velocity distribution function. The density $\rho$ in the
integrand is the dark matter mass density profile in our galaxy.

After integrating out Eq.~(\ref{flux formula 1}) by the solid angle
with the appropriate angular resolution of the detector, we obtain
\begin{eqnarray}
 \frac{d{\cal F}_\gamma(E)}{dE}
 &=&
 9.3 \times 10^{-12}
 \left[{\rm cm}^{-2} {\rm sec}^{-1}{\rm GeV^{-1}}\right]
 \nonumber \\
 &\times&
 \left(
  \frac{100{\rm GeV}}{m}
 \right)^2
 \sum_f
 \frac{dN_f^{(\gamma)}}{dE}
 \left(
  \frac{\left<\sigma v\right>_f}{10^{-27} {\rm cm^3 sec^{-1}}}
 \right)
 \bar{J} \Delta \Omega~.
 \label{flux formula 2}
\end{eqnarray}
The angular resolution of the detector, $\Delta\Omega$, is taken to be
$10^{-3}$ in this paper, which is a typical value for the ACT
detectors. The information of the dark matter density profile is
included in a dimensionless function,
\begin{equation}
 \bar{J}
 =
 \int_{\rm line~of~sight}~\frac{dl(\theta)}{8.5{\rm kpc}}
 \int_{\Delta \Omega}~\frac{d\Omega}{\Delta \Omega}~
 \left(
  \frac{\rho}{0.3~{\rm GeV cm^{-3}}}
 \right)^2~.
 \label{Jbar2}
\end{equation}

We need three quantities for evaluating the gamma ray fluxes; the dark
matter annihilation cross sections to final states $f$
($\langle\sigma v\rangle_f$), the fragmentation functions
($dN^{\gamma}_f/dE$), and the mass density profile of the dark matter
($\rho$). The cross sections are obtained in the previous section. We
discuss the fragmentation functions and the mass density profile in
the following subsections.

\vspace{0.5cm}
\underline{2. Fragmentation function}
\vspace{0.5cm}

\begin{figure}[t]
 \begin{center}
  \includegraphics[height = 5.5cm,clip]{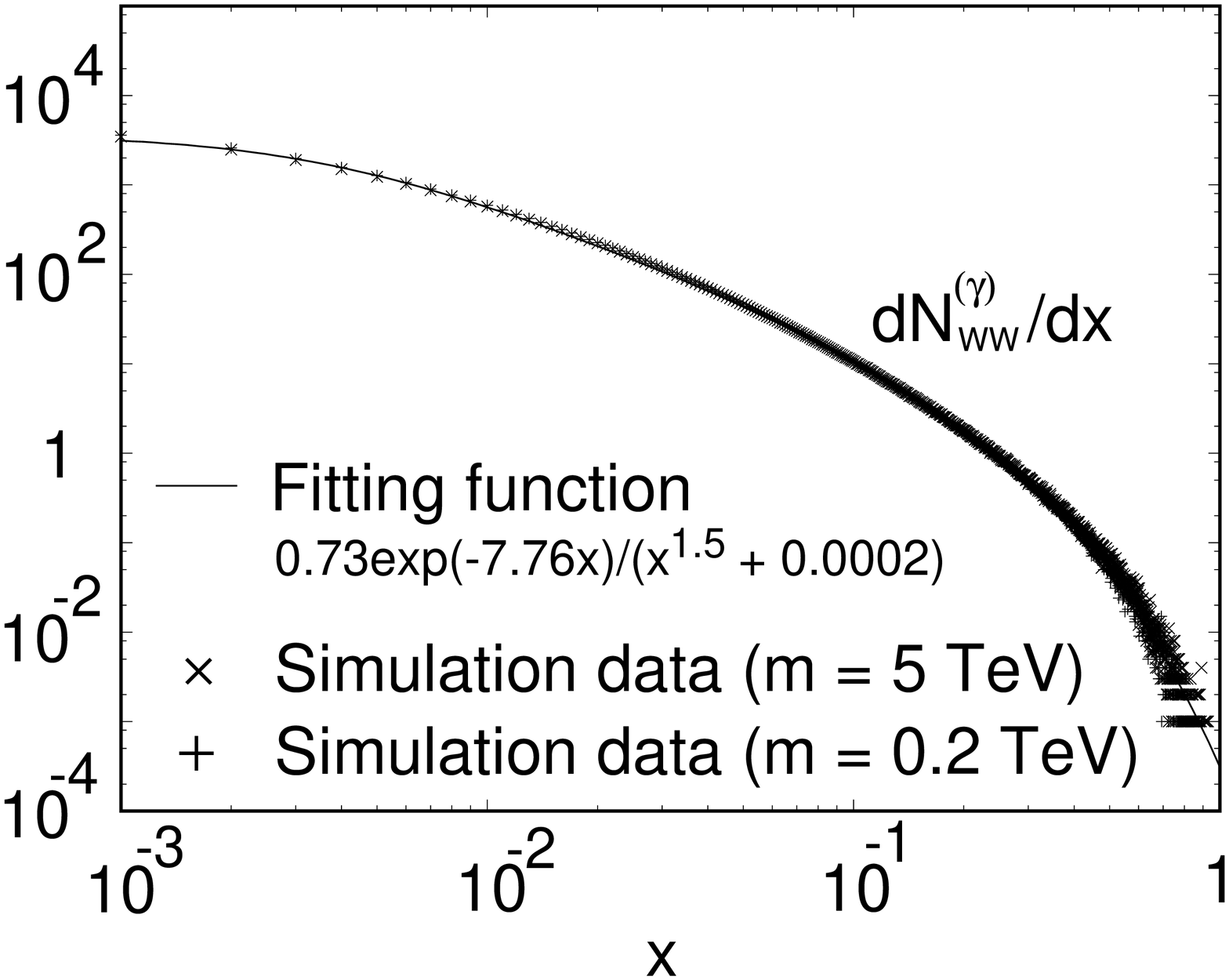}
  ~~
  \includegraphics[height = 5.5cm,clip]{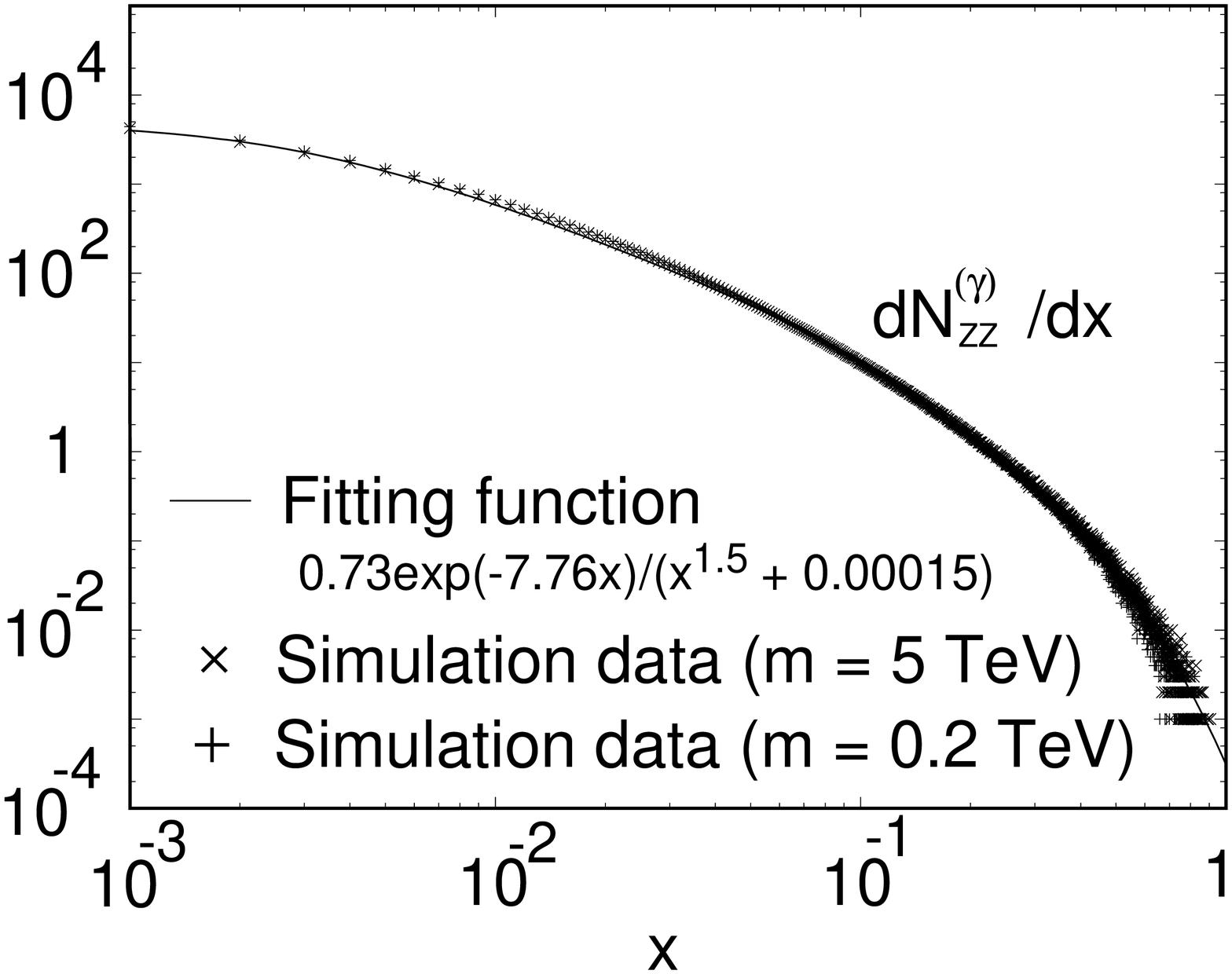}
 \end{center}
 \caption{\small
 Fragmentation functions of $W^+W^-$ (the left figure) and
 $Z^0Z^0$ (the right figure). The cross points are the simulation
 data by HERWIG. The fitting functions are also depicted in
 these figures as solid lines.
 \label{fragmentation function}
 }
\end{figure}

Since we focus on the SU(2)$_L$ non-singlet dark matter, such as the
Wino- or Higgsino-like neutralino in the MSSM, the continuum gamma ray
signal come mainly from the the dark matter annihilation modes into
$W^+W^-$ and $Z^0Z^0$. We thus need two fragmentation functions,
$N_{W^+W^-}^{(\gamma)}$ and $N_{Z^0Z^0}^{(\gamma)}$. We simulate the
photon spectrums from these weak gauge boson states by the HERWIG Monte
Carlo code \cite{herwig}. We derive fitting functions from the
simulated fragmentation functions for $m = 200$ GeV and $m = 5$ TeV by
introducing the scaling variable $x = E/m$.

In Fig.~\ref{fragmentation function}, the HERWIG simulation data for the
fragmentation functions are shown. In these figures, the fitting
functions are also depicted as solid lines. The functions are given by 
\begin{eqnarray}
 \frac{dN_f^{(\gamma)}}{dx}
 =
 \frac{0.73~e^{-7.76x}}{x^{1.5} + c_f}~,
\end{eqnarray}
where the parameter $c_f$ is $2\times 10^{-4}$ for $f = W^+W^-$ and
$1.5\times 10^{-4}$ for $f = Z^0Z^0$.

In the previous studies in Refs.~\cite{Hisano:2002fk}, the cutoff
parameter $c_f$ is 
not introduced in the fitting functions. However, the behaviors of the
simulated fragmentation functions at $x\lsim 10^{-(2 - 3)}$ are more
moderate than the fitting functions with $c_f=0$. In this paper we
consider cases of  heavy EWIMPs $(\sim 10$ TeV) and compare the
predicted gamma ray fluxes with the EGRET data around 1 $\sim$ 10 GeV.
Thus, the effect of non-vanishing $c_f$ is not negligible.

The function $N_{\gamma\gamma}^{(\gamma)}$ for the line gamma ray flux
is simply given by 
\begin{eqnarray}
 \frac{dN_{\gamma\gamma}^{(\gamma)}}{dE}
 =
 2\delta(E - m)~,
\end{eqnarray}
because the dark matter particle is almost at rest in the galactic halo.

\vspace{0.5cm}
\underline{3. Mass density profile of dark matter}
\vspace{0.5cm}

\begin{figure}[t]
 \begin{center}
  \includegraphics[height = 6cm,clip]{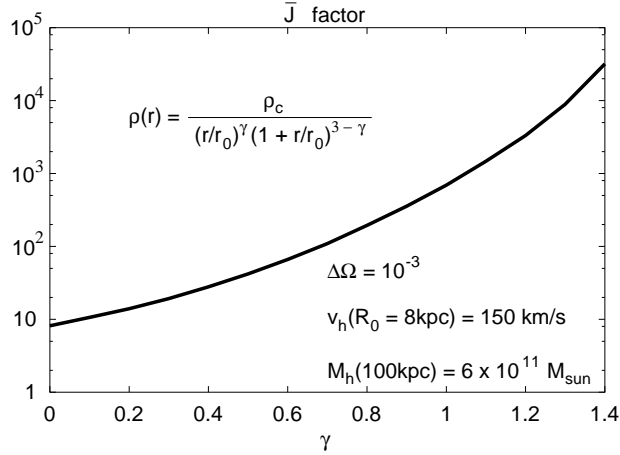}
 \end{center}
 \caption{\small
 Dependence of $\bar{J}$ in Eq.~(\ref{Jbar2}) on $\gamma$ in the dark
 matter halo profile given in Eq.~(\ref{halo model}). We set the other
 parameters $\alpha$ 
 and $\beta$ to be 1 and 3, respectively. The mass of the galaxy $M_h$ and the
 rotational velocity $v_h$, which are used to  determine the parameters
 $\rho_0$ and $a$, are also shown in the figure.
 \label{Jbar}
 }
\end{figure}

The gamma ray flux depends strongly on the dark matter density
profile, because it is proportional to the density squared. Many
$N$-body simulations show that the dark matter halo profiles are given
by a universal spherical functional form,
\begin{eqnarray}
 \rho(r)
 =
 \frac{\rho_0}{(r/a)^\gamma[1 + (r/a)^\alpha]^{(\beta - \gamma)/\alpha}}~,
 \label{halo model}
\end{eqnarray}
where $\alpha$, $\beta$ and $\gamma$ are the model parameters. After
choosing the model parameters, $\rho_0$ and $a$ are uniquely determined
by the mass and the rotational speed of the galaxy.

A famous and frequently used halo model is the Navarro, Frenk and
White (NFW) profile \cite{Navarro:1995iw}, which corresponds to
$(\alpha,\beta,\gamma) = (1,3,1)$ in Eq.~(\ref{halo model}). The
profile is obtained by the numerical $N$-body simulation of point
particles. Recently higher statistical simulations have been
performed, and even higher values of $\gamma$ are obtained. For
example, Moore {\it et.~al.} found the profile corresponding to
$(\alpha,\beta,\gamma) = (1,3,1.4\sim 1.5)$ \cite{Moore:1999nt}. The
halo profiles in all $N$-body simulations have cuspy structures and
the density diverges at $r = 0$. On the other hand, a no-cuspy model
has been used for a long time, and the model parameters for the
King-profile are $(\alpha,\beta,\gamma) = (1,3,0)$. It is argued that
the rotation curve measurements of low surface brightness galaxies
disfavor the cuspy profiles \cite{deblock}, though this disagreement
may be resolved by taking into account the effect of halo triaxiality
\cite{Hayashi:2004vm}. This cusp/core problem is still under debate.

In Fig.~\ref{Jbar}, $\bar{J}$ in Eq.~(\ref{Jbar2})
is shown as a function of the model parameter $\gamma$. Here we fix
$\alpha = 1$ and $\beta = 3$. We use the mass of galaxy interior to 100
kpc ($M_h(100~{\rm kpc}) = 6\times 10^{11}M_{\rm sun}$ with $M_{\rm
sun}$ the solar mass) and the rotational speed at $r = 8$ kpc
($v_h(8~{\rm kpc}) = 150$ km/s) for the determination of $\rho_0$ and $a$ in
Eq.~(\ref{halo model}). In the figure $\bar{J}$ is sensitive to the
value of $\gamma$. In the following, we use a moderate value
$\bar{J}=500$, which is typical for the NFW profile.

\vspace{0.5cm}
\underline{4. Line gamma ray flux from the galactic center}
\vspace{0.5cm}

\begin{figure}[t]
 \begin{center}
  \includegraphics[height = 4.9cm,clip]{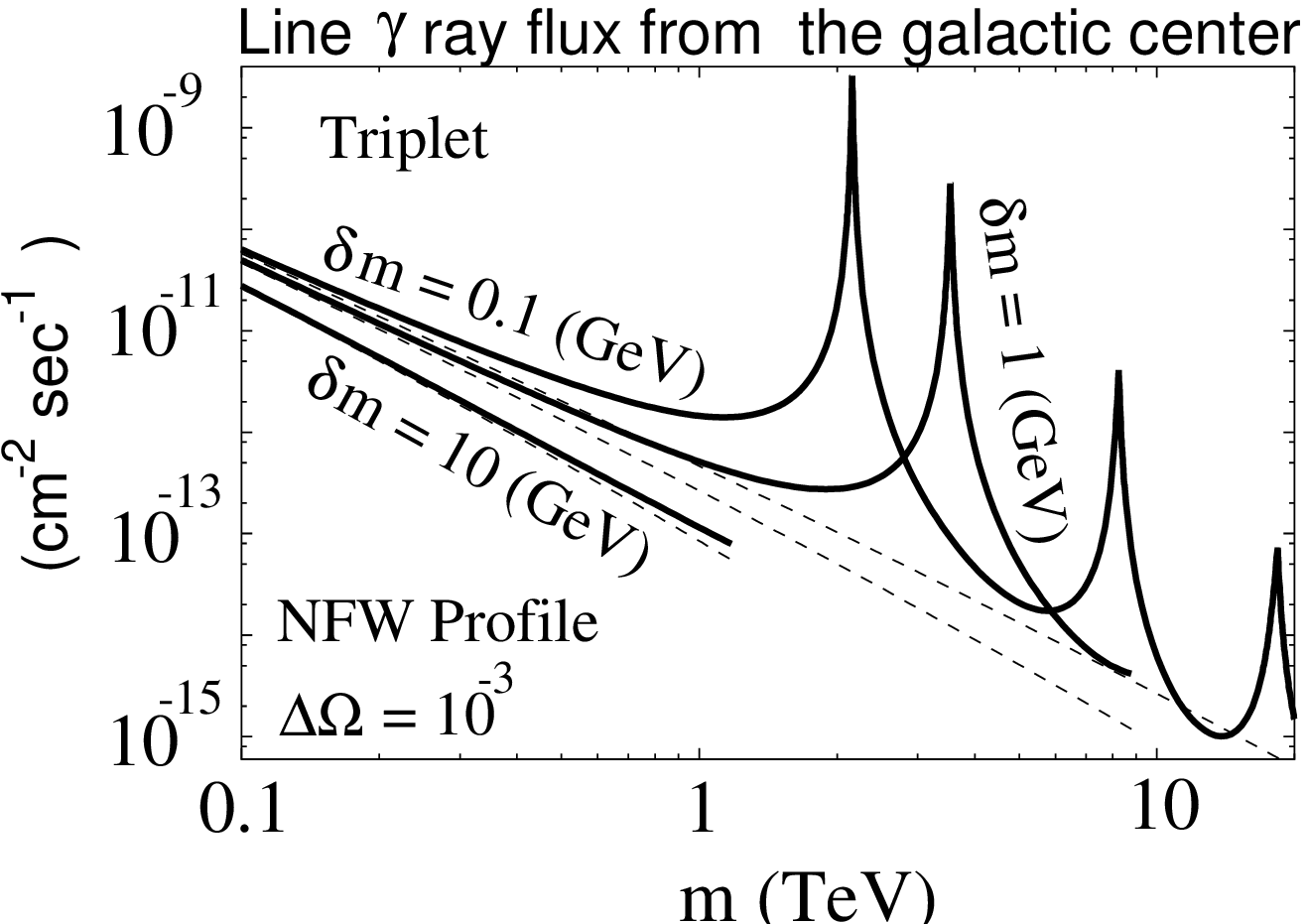}
  ~~~
  \includegraphics[height = 4.9cm,clip]{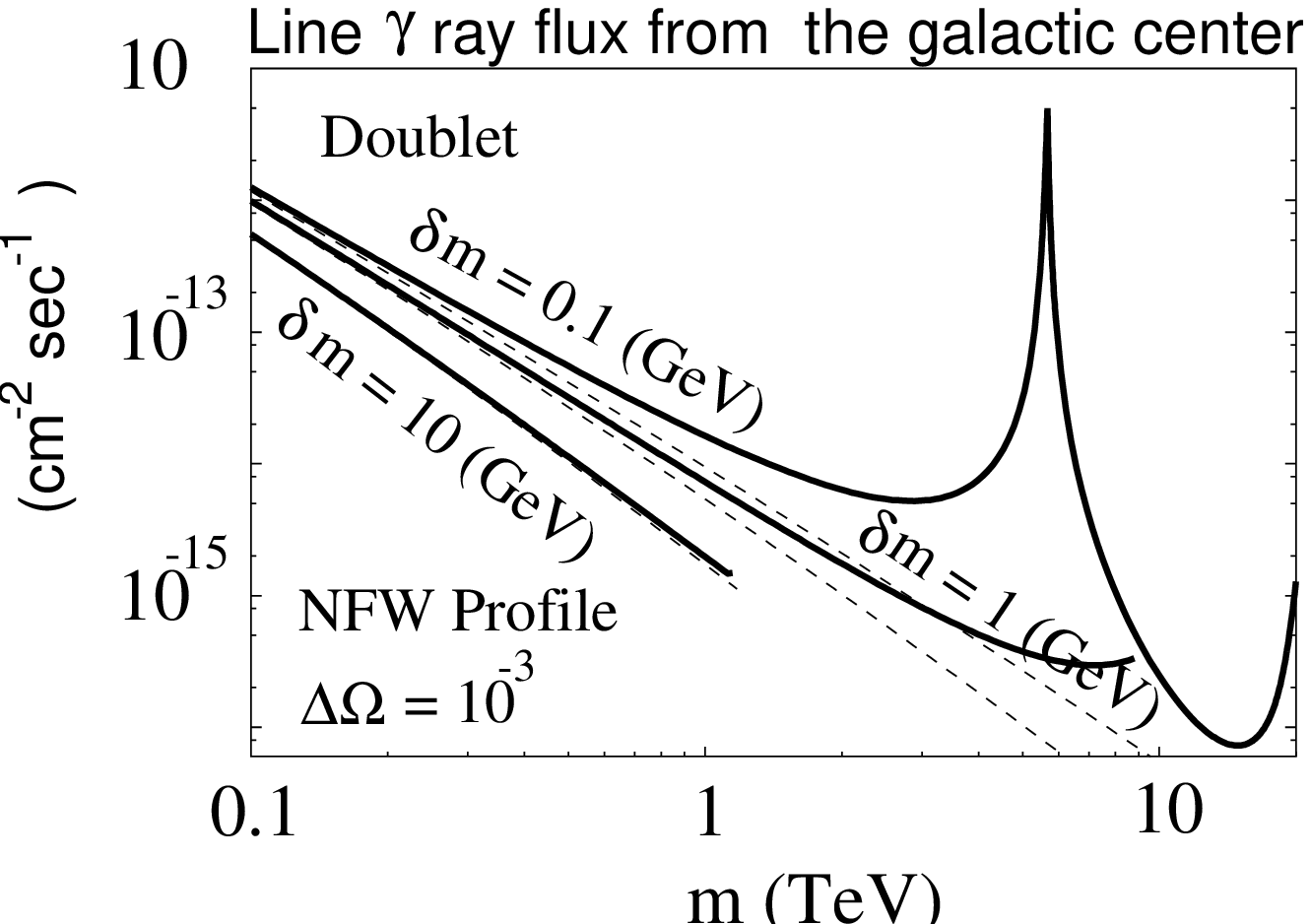}
 \end{center}
 \caption{\small
 Line gamma ray flux from the galactic center in cases of the EWIMP 
dark matters. The left (right) figure is for the triplet (doublet) EWIMP.
We take the average velocity of the dark matter
 $v/c = 10^{-3}$, $\bar{J} = 500$, which corresponds to the NFW profile,
 and $\Delta\Omega = 10^{-3}$. The leading-order cross sections
 in perturbation are also shown as broken lines.
 \label{line gamma 1} }
\end{figure}

We discuss the line gamma ray flux from the galactic center
due to the EWIMP dark matter annihilation by using the cross sections
derived in the previous section.

\begin{figure}[p]
 \begin{center}
  \includegraphics[height = 9cm,clip]{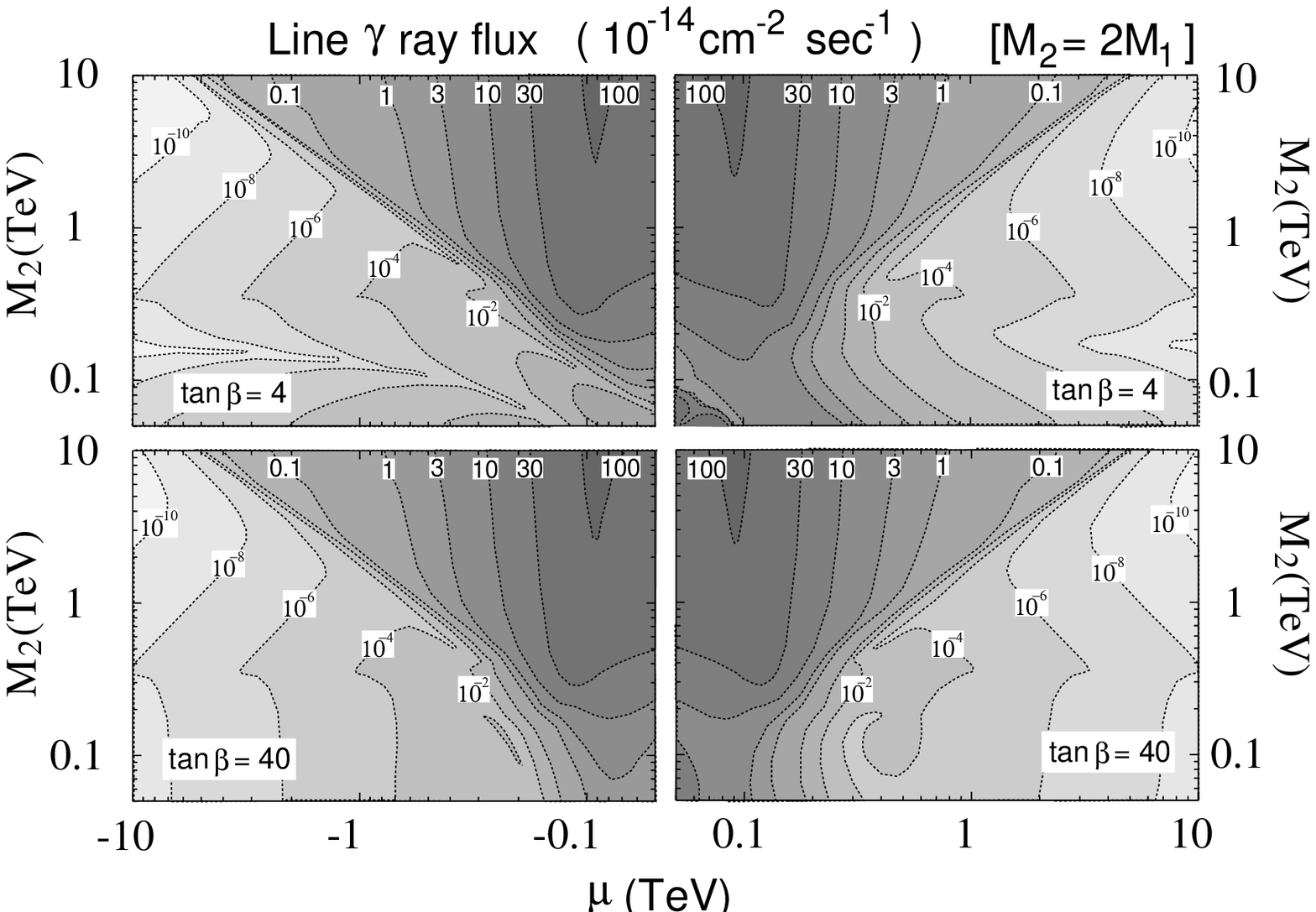}
  \\
  \rule{1cm}{0cm}
  \\
  \includegraphics[height = 9cm,clip]{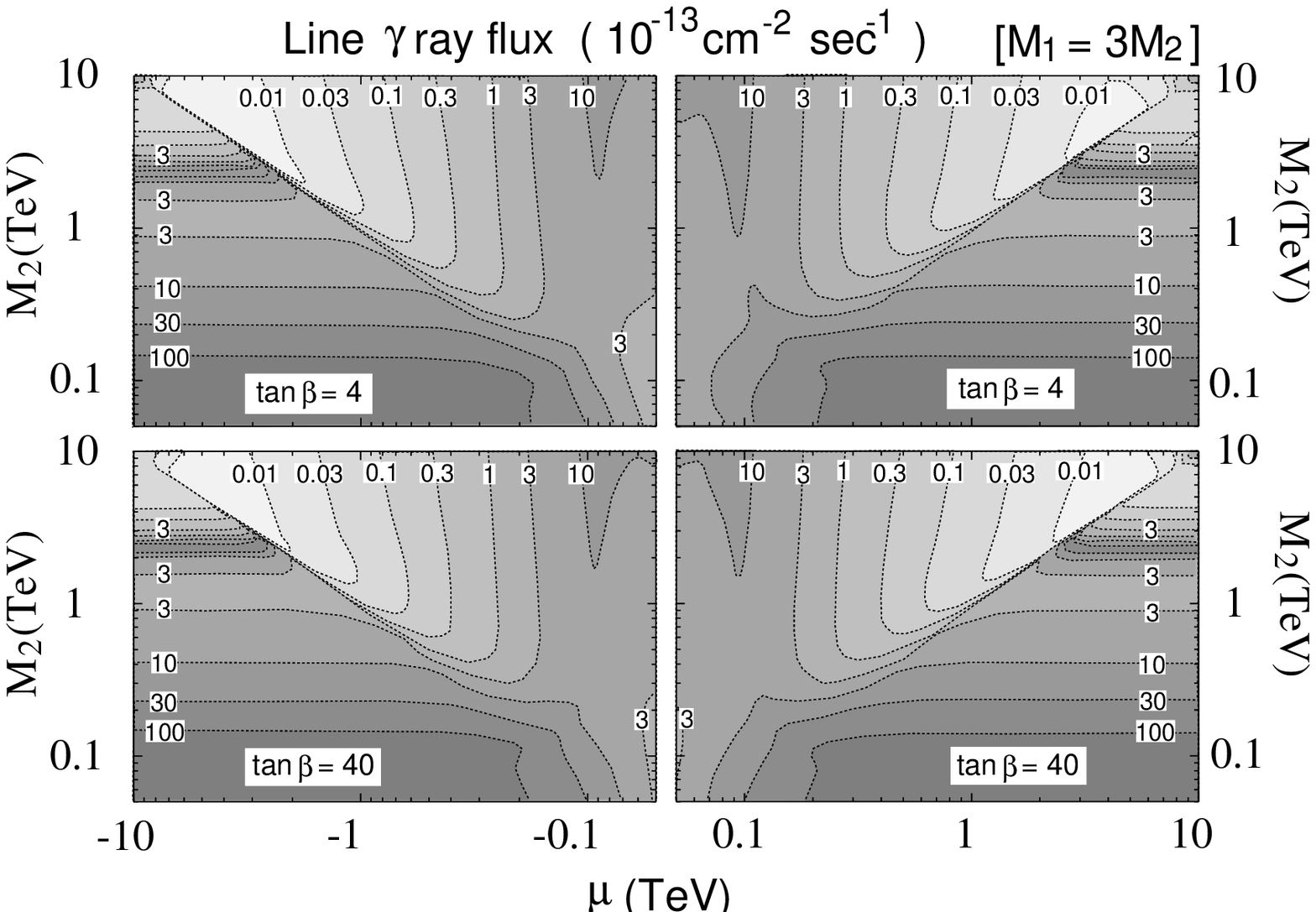}
 \end{center}
 \caption{\small
 Contour maps of the line gamma ray flux in ($M_2,\mu$) planes with
 $\tan\beta = 4, 40$ in the MSSM. $M_1 = 0.5M_2$
 is assumed in the four top figures while  $M_1 = 3M_2$ is
 assumed in  the four bottom figures. These figures are depicted by using
 the  NFW profile ($\bar{J} = 500$) and the angular resolution,
 $\Delta\Omega = 10^{-3}$.
 \label{ScanLine}
 }
\end{figure}

In Fig.~\ref{line gamma 1}, we show the line gamma fluxes from the
galactic center in the cases of the triplet and doublet EWIMP dark
matters. Here, we take the mass difference between 
the EWIMP and the charged partner as $\delta m=0.1,1,10$~GeV. 
For the doublet EWIMP, $\delta m_N = 2\delta m$ is assumed.
We also plot
the flux obtained from the leading-order cross sections in 
perturbation by broken lines for comparison. In Fig.~\ref{ScanLine}, 
we show the contour maps of the line gamma ray flux in the
MSSM. The range of the MSSM parameters is the same as that
for Figs.~\ref{CSI} and \ref{CSII} in the previous section.

The large ACT detectors have high sensitivity for TeV-scale gamma
rays. For example, MAGIC \cite{Baixeras:2003xr} and VERITAS
\cite{Weekes:1997np} in the  northern hemisphere might
reach  $10^{-14}$cm$^{-2}$s$^{-1}$ at the TeV scale while CANGAROO III
\cite{Tsuchiya:2004wv} and HESS \cite{Hinton:2004eu} in the southern
hemisphere might reach $10^{-13}$cm$^{-2}$s$^{-1}$. From
Figs.~\ref{line gamma 1} and \ref{ScanLine}, it is found that these
ACT detectors may cover broad regions in the parameter space.

It has been known that the line gamma ray signal is sensitive to the
heavier dark matter with the mass of the order of TeV, because the
annihilation cross section at one-loop level is not suppressed by the
dark matter mass if the dark matter has the SU(2)$_L$ charge. Our
studies reveal importance of the non-perturbative effects on the
EWIMP annihilation cross section. After including the effects, the
sensitivity of the line gamma ray signal to the heavier EWIMP dark
matter is enhanced furthermore.

\vspace{0.5cm}
\underline{5. Continuum gamma ray flux from the galactic center}
\vspace{0.5cm}

\begin{figure}[t]
 \begin{center} \includegraphics[height = 5cm,clip]{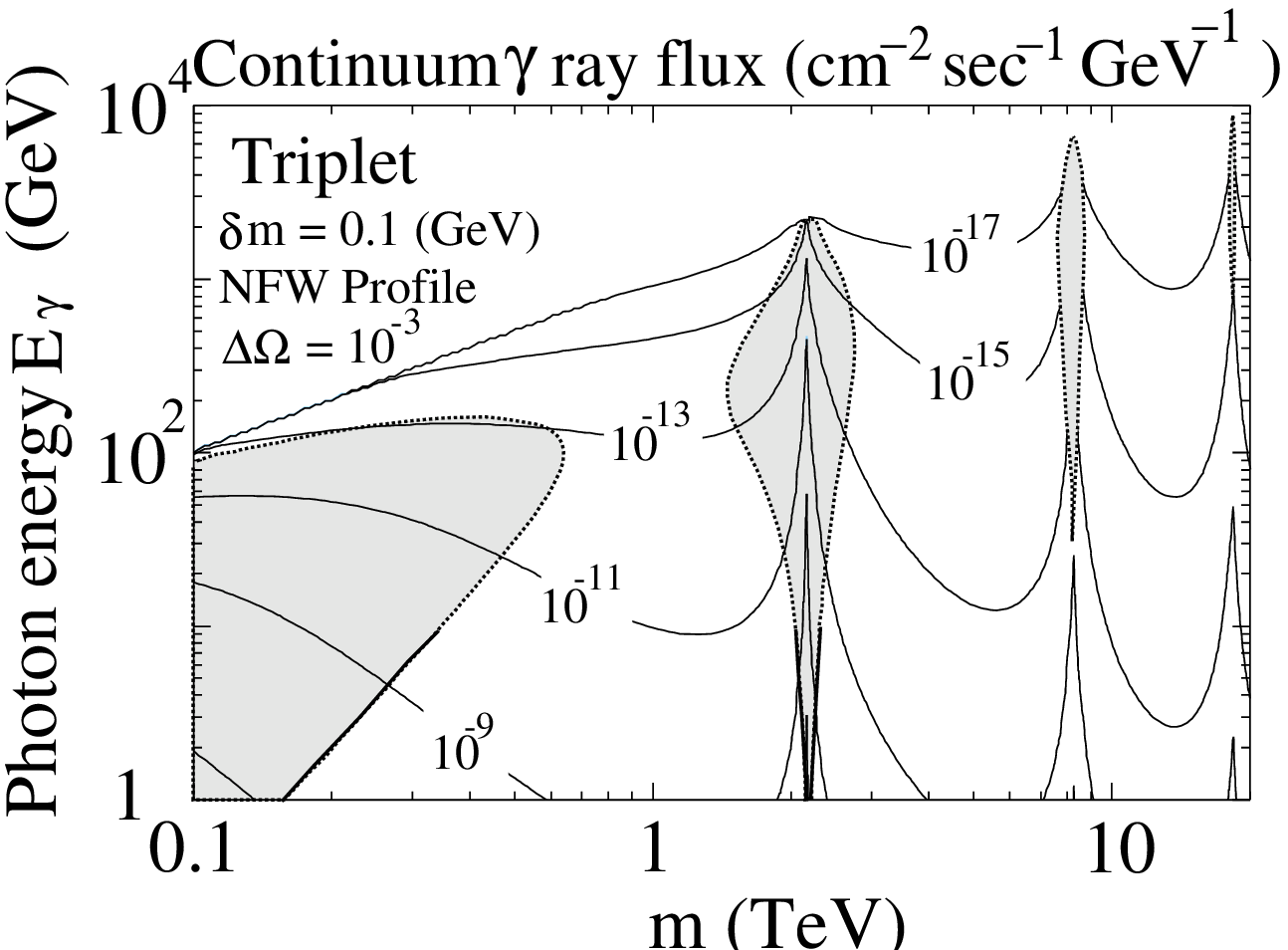} ~~~
 \includegraphics[height = 5cm,clip]{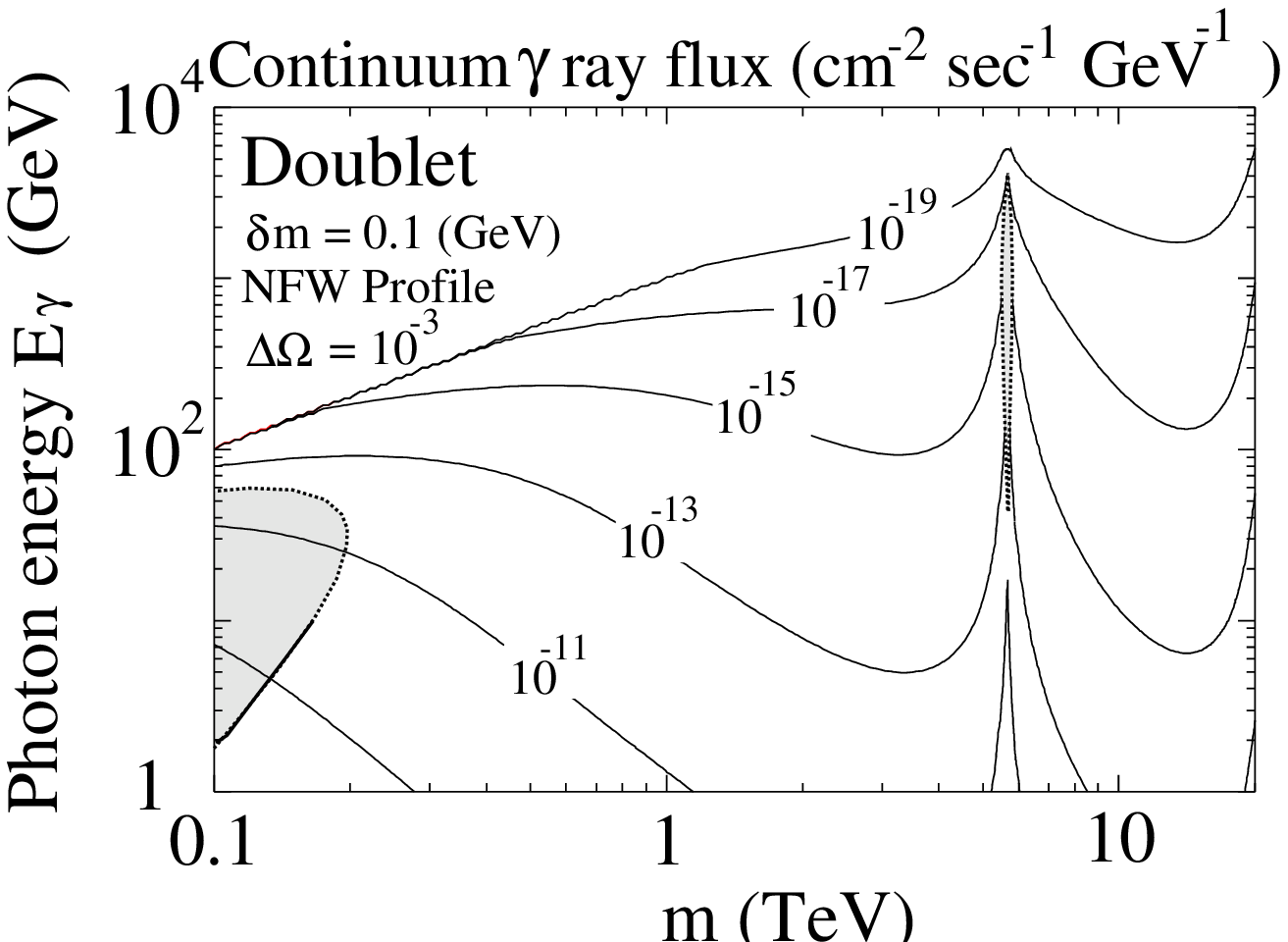} \end{center}
 \caption{\small Contour plots of the continuum gamma ray flux from
 the galactic center in the unit of cm$^{-2}$sec$^{-1}$GeV$^{-1}$. The
 left (right) plot corresponds to the triplet (doublet) EWIMP case.
 Here we take $\delta m = 0.1$ GeV, and other parameters are the same
 as those in Fig.~\ref{line gamma 1}. The shaded regions correspond to
 (Signal)/(Background) $>$ 1.  \label{cont gamma 1} }
\end{figure}

In addition to the line gamma rays, the EWIMP annihilation produces
the continuum gamma rays, which come mainly from the decay of $\pi^0$
in the fragmentation of the final state particles. Though the number
of photons in the continuum gamma rays is expected to be higher than
that in the line gamma rays, its spectrum may lack the distinctive
feature. Since the flux of the diffused gamma ray background,
especially from the galactic center, is not well known, it is
difficult to extract the annihilation signal ($S$) from the background
($B$) as far as $S<B$.  The EGRET has observed the diffused
gamma ray emission from the galactic center up to about 10 GeV
\cite{Hunger:1997we}. If the power law fall-off of the energy for the
diffused gamma ray flux $\Psi_{BG}(E)$ is assumed, the background $B$
is evaluated from the EGRET result as \cite{Bergstrom:1997fj},
\begin{eqnarray}
\frac{d \Psi_{BG}(E)}{dE} 
=9.1\times 10^{-5}
[{\rm cm^{-2} sec^{-1} GeV^{-1}}]
\times
\left(\frac{E}{1{\rm GeV}}\right)^{-2.7}\Delta \Omega.
\label{egret}
\end{eqnarray}

In Fig.~\ref{cont gamma 1} we show the contour plots of the continuum
gamma ray fluxes 
from the galactic center for the triplet and doublet EWIMPs. Here we
fix $\bar{J}=500$, $\Delta \Omega=10^{-3}$ and $\delta m=0.1$ GeV.
The shaded regions correspond to $S>B$, in which $B$ is given by
Eq.~(\ref{egret}). From this figure, it is found that even small
regions around the resonances, in addition to areas with $m\sim100$
GeV, are already constrained from the EGRET measurement of the gamma
ray flux with the energy $1\sim 10$ GeV.  

\begin{figure}[p]
 \begin{center}
  \includegraphics[height = 9cm,clip]{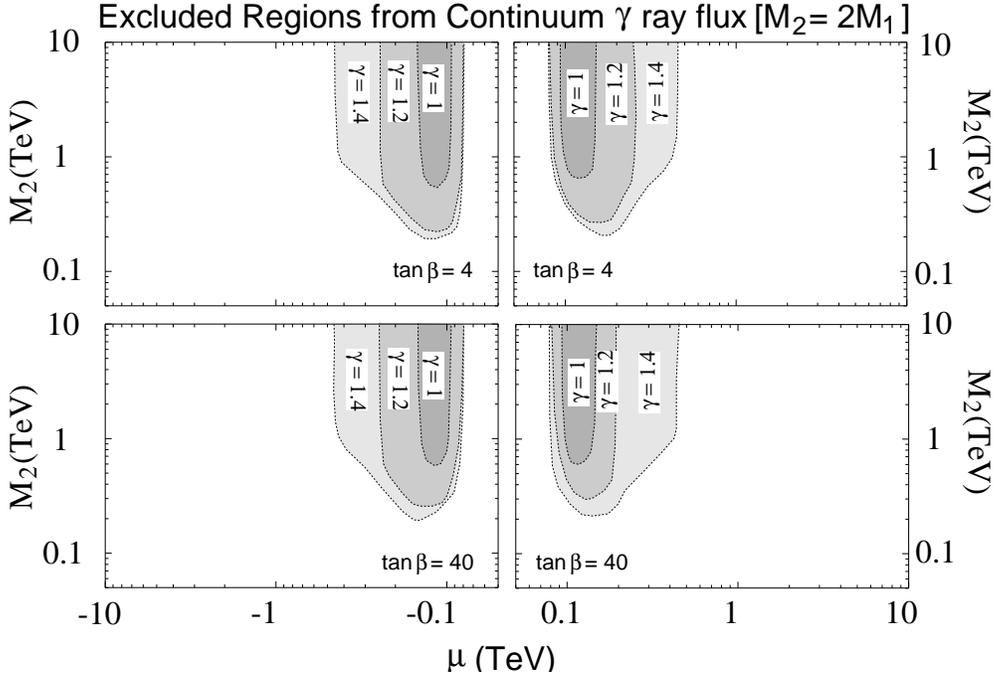}
  \\
  \rule{1cm}{0cm}
  \\
  \includegraphics[height = 9cm,clip]{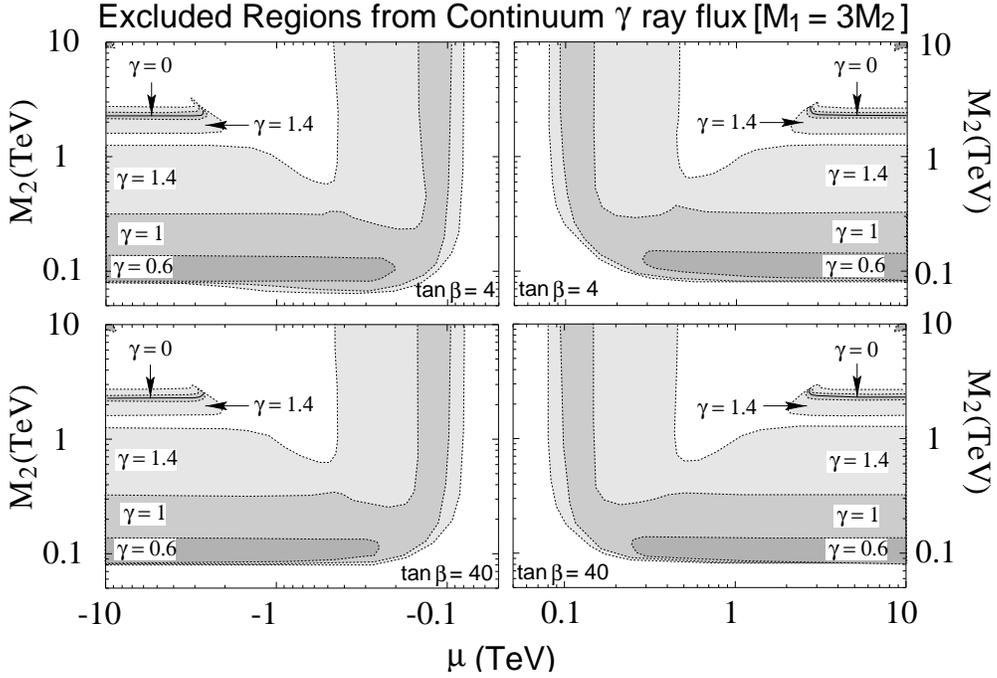}
 \end{center}
 \caption{\small
Contour maps of the excluded region by the EGRET measurement of the
gamma ray flux from the galactic center in ($M_2,\mu$) planes in the
MSSM for different dark matter profiles. $M_1 = 0.5M_2$ is assumed in
the four top figures while it is $M_1 = 3M_2$ in the four bottom
figures. $\tan\beta = 4, 40$. These figures are depicted by using the
$\bar{J}$ factors for $\gamma = 0, 0.6, 1, 1.4$. We also set the
angular resolution to be $\Delta\Omega = 10^{-3}$.
 \label{ScanCont}
 }
\end{figure}

In Fig.~\ref{ScanCont} regions excluded by the EGRET observation is
shown in 
$(M_2,\mu)$ planes assuming the MSSM. The MSSM parameters are the same
as in Fig.~\ref{ScanLine}. In this figure we use $\bar{J}$s with
$\gamma=0$, 0.6, 
1 and 1.4. The excluded regions become broader for larger $\gamma$.

In the near future the GLAST satellite \cite{Morselli:1997rk} may observe
gamma rays with energies in the range 1 GeV$\lsim E \lsim$ 300 GeV if
the flux is larger than about $10^{-10} {\rm cm^{-2}sec^{-1}}$. The
ACT and GLAST detectors will constrain broader regions in the
parameter space with $S>B$, or will find the signature of the dark
matter.

\vspace{1.0cm}
\lromn 7 \hspace{0.2cm}
{\bf Summary}
\vspace{0.5cm}

In this paper we have calculated the pair annihilation cross sections of
the EWIMP dark matters, which have an SU(2)$_L$ charge of the standard
model gauge group. The leading-order calculation of the cross sections
in perturbation is no longer valid when the mass is heavy compared
to the weak gauge  bosons due to the threshold singularity coming from
the mass degeneracy between the EWIMP and its SU(2)$_L$ partner(s). The
problem have been known for a while for the cases of the Wino and
Higgsino-like dark matters in the MSSM. We have developed a
method to take in the singularity and obtain the precise annihilation
cross sections.

We find that if the mass of the EWIMP dark matter is larger than about 1
TeV, the attractive Yukawa potentials induced by the weak gauge boson exchanges
have significant effects on the annihilation processes and the cross
sections are enhanced by several orders of magnitude due to the zero-energy
resonances under the potentials. As a result, the gamma ray flux 
from the galactic center due to the EWIMP annihilation is enhanced
compared to the leading-order calculation in perturbation. The line
gamma ray flux exceeds the typical sensitivity of the large ACT detectors such 
as CANGAROO III, HESS, VERITAS and MAGIC, $10^{-(13-14)}$
cm$^{^-2}$sec$^{-1}$, in the wide range of the MSSM parameters. We also
calculated the continuum gamma ray flux from the EWIMP dark matter
annihilation. The MSSM parameter space which is not consistent with the
EGRET observation of the gamma rays with 1 GeV $\lsim E_\gamma \lsim$ 10
GeV is increased by the non-perturbative effects. The
non-perturbative effect would also enhance anti-proton and positron
fluxes from the dark matter annihilation. These will be discussed
elsewhere \cite{Hisano:2005}. 

Current observations of TeV-scale gamma rays from the galactic center by
the CANGAROO and HESS disagree each other in the spectrum. Once it
is converged, the result may be used to constrain the EWIMP dark matter.

\vspace{1cm}

\underline{Acknowledgments}
\vspace{0.5cm}

This work is supported in part by the Grant-in-Aid for Science
Research, Ministry of Education, Science and Culture, Japan
(No.~13135207 and 14046225 for JH
and No.~14540260, 14046210 and 16081207 for MMN).
MMN is also supported in part by a Grant-in-Aid 
for the 21st Century COE "Center for Diversity and Universality in
Physics".

\vspace{0.5cm}


\end{document}